\documentclass[superscriptaddress,dvipsnames,nofootinbib,amsmath,amssymb,prd,notitlepage]{revtex4-1}

\usepackage{subfig}
\usepackage[a4paper, left=20mm,right=20mm,top=20mm,bottom=20mm]{geometry}
\usepackage[titletoc,toc,title]{appendix}

\usepackage{mathtools}
\usepackage{tensor}
\usepackage{amsmath, amsfonts,bm}
\usepackage{derivative}
\usepackage{bm}
\usepackage[utf8]{inputenc}
\usepackage[normalem]{ulem}

\usepackage{suffix}

\usepackage{amsmath, amsfonts,bm}
\usepackage{xcolor}
\usepackage{graphicx}
\definecolor{darkerred}{rgb}{0.9, 0.17, 0.31}
\usepackage{subfig}

\usepackage[colorlinks=true,hyperfootnotes=true, linkcolor=darkerred, citecolor=cyan, urlcolor=magenta]{hyperref}

\newcommand\mathcomma{\,,}
\newcommand\mathperiod{\,.}

\DeclareMathAlphabet{\mathup}{OT1}{\familydefault}{m}{n}
\newcommand\imagi{\mathup{i}}
\newcommand\expe{\mathup{e}}

\def\dd{\mathrm{d}}

\newcommand\covd{\nabla}

\newcommand{\be}{\begin{equation}} 
\newcommand{\ee}{\end{equation}}

\usepackage{array}
\newcommand{\PreserveBackslash}[1]{\let\temp=\\#1\let\\=\temp}
\newcolumntype{C}[1]{>{\PreserveBackslash\centering}p{#1}}
\newcolumntype{R}[1]{>{\PreserveBackslash\raggedleft}p{#1}}
\newcolumntype{L}[1]{>{\PreserveBackslash\raggedright}p{#1}}

\begin{document}

\title{Dissecting kinetically coupled quintessence:\\ phenomenology and observational tests}

\author{Elsa M. Teixeira}
\email{emcteixeira1@sheffield.ac.uk}
\affiliation{School of Mathematics and Statistics, University of Sheffield, Hounsfield Road, Sheffield S3 7RH, United Kingdom}
\author{Bruno J. Barros}
\email{cstbru002@myuct.ac.za}
\affiliation{Cosmology and Gravity Group, Department of Mathematics and Applied Mathematics, University of Cape Town, Rondebosch 7700, Cape Town, South Africa}
\author{Vasco M. C. Ferreira}
\email{vasco.ferreira@astro.up.pt}
\affiliation{Instituto de Astrof\'{\i}sica e Ci\^encias do Espa{\c c}o, Universidade do Porto, CAUP, Rua das Estrelas, PT4150-762 Porto, Portugal}
\author{Noemi Frusciante}
\email{noemi.frusciante@unina.it}
\affiliation{Dipartimento di Fisica ``E. Pancini", Universit\`a degli Studi  di Napoli  ``Federico II", Compl. Univ. di Monte S. Angelo, Edificio G, Via Cinthia, I-80126, Napoli, Italy}

\begin{abstract}
We investigate an interacting dark energy model which allows for the kinetic term of the scalar field to couple to dark matter \textit{via} a power-law interaction. The model is characterised by scaling solutions at early times,  which are of high interest to alleviate the coincidence problem, followed by a period of accelerated expansion.  We discuss the phenomenology of the background evolution and of the linear scalar perturbations and we identify measurable signatures of the coupling in the dark sector on the cosmic microwave background, the lensing potential auto-correlation and the matter power spectra. We also perform a parameter estimation analysis using data of cosmic microwave background temperature, polarisation and lensing, baryonic acoustic oscillations  and supernovae. We find that the strength of the coupling between the dark sectors, regulated by the parameter $\alpha$, is constrained to be of order $10^{-4}$. A model selection analysis does not reveal a statistical preference between $\Lambda$CDM and the Kinetic model.
\end{abstract}

\maketitle

{\hypersetup{linkcolor=black}
\tableofcontents
}

\pagebreak

%-----------------
\section{Introduction}\label{sec:intro}
%-----------------

The existence of dark energy (DE) and dark matter (DM) is supported by multiple cosmological observations, though their nature still remains unknown.
The former is postulated as a repulsive force acting on the largest scales, needed to explain the late time cosmic acceleration, whereas the latter is a non-baryonic matter component, responsible for the  formation  and evolution of large scale structures in the Universe. The Standard Model of Cosmology, known as $\Lambda$-cold dark matter ($\Lambda$CDM), is based on General Relativity (GR) and includes a cosmological constant, $\Lambda$, as the simplest model of DE and a cold dark matter (CDM) component as weakly interacting non-relativistic particles. In this base scenario, it is assumed that the two dark components do not directly couple to each other. Although it provides a fairly accurate description of the Universe, there are some unexplained theoretical and observational conundrums that indirectly affect the $\Lambda$CDM model \cite{Weinberg:1988cp, Weinberg:2000yb, Martin:2012bt}. Such is the case of the Cosmological Constant problem or the need for a primordial inflationary period. Observational tensions, if not stemming from systematics, pose an additional challenge \cite{Abdalla:2022yfr} namely concerning the mismatch in the estimation of the values of the Hubble constant, $H_0$ \cite{Planck:2018vyg, Riess:2019cxk, Wong:2019kwg, Riess:2020fzl, Pesce:2020xfe}, and the amplitude of the matter power spectrum at present time, $\sigma_8$ \cite{Heymans:2020gsg, DiValentino:2020vvd, KiDS:2020suj}, when using high- and low-redshift data from different surveys. These shortcomings might signal the need to go beyond the vanilla $\Lambda$CDM model \cite{CANTATA:2021ktz}. 

Promoting DE to a dynamical scalar field is an enticing approach to extend $\Lambda$CDM and still achieve the late-time accelerated expansion. Recent experimental advancements in particle physics have lead to the detection of a Higgs-like particle \cite{ATLAS:2012yve, CMS:2012qbp} and scalar fields also comprise the most promising proposal to solve the early Universe trinity puzzle ({\it i.e.} the horizon, flatness, and magnetic-monopole problems) \cite{Guth:1980zm, Linde:1981mu, Starobinsky:1982ee}. The quintessence model \cite{Wetterich:1994bg, Caldwell:1997ii, Chiba:1999wt} was the first attempt to include a scalar degree of freedom, $\phi$, portraying a time-varying DE component with dynamics assigned by the form of the  potential, $V(\phi)$, and its kinetic term, ${X=-\partial_{\mu}\phi\partial^{\mu}\phi}/2$.  
 
In particular, it should resemble the cosmological constant at late times, that is, its negative pressure must have a magnitude close to its energy density, $p_{\phi} \approx - \rho_{\phi}$, while not revealing effective clustering properties at small scales.
One  appealing feature of this theory (or, more in general, of scalar tensor theories) lies in obtaining, under particular conditions, scaling solutions \cite{Wetterich:1994bg,Copeland:1997et,Ferreira:1997hj,Liddle:1998xm,Barreiro:1999zs,Amendola:1999er,Guo:2003rs,Guo:2003eu,Chimento:2003iea,Tsujikawa:2004dp,Piazza:2004df,Pettorino:2005pv,Amendola:2006qi,Ohashi:2009xw,Gomes:2013ema,Chiba:2014sda,Amendola:2014kwa,Albuquerque:2018ymr,Frusciante:2018aew,Amendola:2018ltt,Barros:2019rdv,Albuquerque:2021grl,Abdalla:2022yfr,Pace:2022nee}. These are characterised by a constant ratio between the energy density of the matter components and that of the scalar field. In this case the DE contribution remains hidden throughout the radiation and matter domination eras, despite allowing the energy density of the scalar field to be of the same order of magnitude as these components. This mechanism is relevant in addressing the cosmic coincidence problem \cite{Zlatev:1998tr,Velten:2014nra}, namely why the magnitude of the energy densities for matter and DE are comparable at present, while still preserving compatibility with the energy scale associated with particle physics. Accordingly, here we will focus on a specific model that already revealed to feature scaling solutions \cite{Barros:2019rdv}.

In this work we are interested in exploring a setting in which the scaling regime is achieved  through a ``fifth-force" acting on DM particles, induced by a quintessence field. An effective field theory formulation of such phenomenological interaction can be set at the level of the action and provides a fully covariant way to construct theoretically viable models  \cite{Tamanini:2015iia}, thus avoiding the propagation of unphysical modes on large scales \cite{Valiviita:2008iv}. 
One such approach consists of considering the presence of a field dependent function $f\left(\phi\right)$ multiplying the CDM Lagrangian, $\mathcal{L}_c$, in the total action, that is, a coupling of the form $f(\phi)\mathcal{L}_c$ \cite{Koivisto:2005nr}. Recently, this formulation was generalised to accommodate interactions of the matter sector with the kinetic term of the scalar as well, through a functional form $f\left(\phi\,,X\right)\mathcal{L}_c$ \cite{Barros:2019rdv}. Lagrangian-based models have been further explored in the context of the Schutz-Sorkin action \cite{Schutz:1970my, Schutz:1977df, Brown:1992kc}, allowing for the inclusion of interaction terms depending on single derivatives of the scalar field in the action for CDM \cite{Pourtsidou:2013nha, Boehmer:2015kta, Boehmer:2015sha}. Along similar lines, in Ref.~\cite{Kase:2019veo} the energy exchange is achieved {\it via} two terms of the form $f_1(\phi,X)\rho_c(n_c)$ and $f_2(n_c,\phi,X)J_c^\mu\partial_\mu\phi$, where $\rho_c$ and $n_c$ are the energy density and number density of CDM, respectively, and $J_c^\mu$ is a vector field related to the CDM four-velocity \cite{Kase:2019veo}. In the presence of a $f(\phi)$-coupling, scaling solutions have been shown to exist for quintessence with an exponential potential \cite{Amendola:1999qq,Amendola:1999er}. Likewise, general forms  of the Lagrangian allowing for scaling behaviour given either a constant or field-dependent interaction, have been derived for k-essence \cite{Piazza:2004df,Tsujikawa:2004dp,Tsujikawa:2006mw,Amendola:2006qi} and scalar-tensor theories such as Horndeski \cite{Gomes:2013ema,Gomes:2015dhl,Amendola:2018ltt,Frusciante:2018aew} and quadratic-order degenerate higher-order scalar-tensor theory \cite{Frusciante:2018tvu}.

Setting cosmological constraints on the interaction between a scalar field and DM has been the subject of many investigations \cite{Amendola:2003eq,Pettorino:2008ez,Bean:2008ac,Pettorino:2012ts,Pettorino:2013oxa,Xia:2013nua,Planck:2015bue,vandeBruck:2016hpz,Pourtsidou:2016ico,VanDeBruck:2017mua,Barros:2018efl,Agrawal:2019dlm,Gomez-Valent:2020mqn,Pan:2020zza,daFonseca:2021imp,Archidiacono:2022iuu}. A well tested class of proposals is the coupled DE model in which DM particles interact with the scalar field due to a $\phi$-dependent mass, characterised by a constant coupling strength $\beta$. This parameter has been constrained to be $\beta=0.036\pm 0.016$ (\textit{Planck}13 + WMAP + baryon acoustic oscillations (BAO)), deviating from the vanishing interaction case at $2.2\sigma$, and $\beta=0.066 \pm 0.018$ when including polarisation, with increasing significance at 3.6$\sigma$ \cite{Pettorino:2013oxa}. 
Similar results were reported lately by the \textit{Planck} collaboration, also showing a tension at $\sim 2.5\sigma$ with $\Lambda$CDM when  \textit{Planck}15 + BAO + Supernovae Ia + $H_0$ data are considered \cite{Planck:2015bue}, and in Refs.~\cite{vandeBruck:2016hpz,Barros:2018efl,Gomez-Valent:2020mqn} resorting to more recent data sets. 
Additionally it has been realised that such constant coupling can remove the $\sigma_8$ tension  if the background is assumed to be identical to the $\Lambda$CDM one \cite{Barros:2018efl}.
Moreover in Ref.~\cite{Bean:2008ac} the authors provided cosmological bounds for a variety of models, which differ from each other through the form of the nontrivial coupling between the DM and the quintessence field. The strength of the coupling was constrained to be less than 7\% of the coupling to gravity.

Let us remark that non-minimal couplings of the DE field to other matter components have also been explored, \textit{e.g.} to massive neutrinos \cite{Afshordi:2005ym, Brookfield:2005td}, to baryons \cite{Aviles:2010ui} or to the electromagnetic field \cite{Carroll:1998zi, Chiba:2001er}. A universal coupling has also been investigated and its magnitude is tightly constrained through Solar System experiments \cite{Hui:2009kc,Creminelli:2013nua}.
Therefore such couplings are often chosen to be \textit{minimal}, \textit{i.e.} there is no additional coupling of the matter fields to the scalar curvature, thus motivating the choice of a direct coupling between the dark species only.

In this work we explore the model presented in Ref.~\cite{Barros:2019rdv}, in which a purely kinetic coupling between the quintessence field and DM is considered.
This coupling is expressed in terms of a power law interaction function, $f\propto X^{\alpha}$, with $\alpha$ being a constant parameter quantifying the strength of the interaction. Hereafter this will be referred to as the Kinetic model. At a more fundamental level, the low-energy limit of a scalar field theory with a shift symmetry only allows for kinetic couplings to matter \cite{Brax:2016kin}, where the scalar field is identified as the Goldstone mode of the broken symmetry. Although in the literature it is much more natural to consider a universal coupling, such as in dilaton gravity \cite{Damour:1994zq}, it is possible to construct a specific (non-universal) interaction with an individual matter source \cite{Damour:1990tw} or it can even naturally emerge in an effective description of a fundamental theory, such as Type II string theory \cite{Koivisto:2013fta}. The toy model considered in this present work also allows for scaling solutions at early times \cite{Barros:2019rdv}, already found to be fruitful to tackle the cosmic coincidence problem. The specific kinetic power law coupling here assumed was employed in the literature to couple quintessence to electromagnetism \cite{Barros:2022kpo} inducing a time variation on the fine-structure constant. The authors showed that the theory encapsulates a plethora of new analytical coupled solutions motivated by the dark energy kinematics. We remark that kinetically coupled models have never been fully explored in terms of theoretical predictions  at linear order in perturbations and, as such, cosmological bounds on the parameters are not present in literature. In this work we present such kind of analysis for the first time, by comparing the theoretical predictions  to the $\Lambda$CDM model for the temperature-temperature (TT) power spectrum, lensing potential auto-correlation power spectrum and matter power spectrum. These are then used to provide cosmological constraints by means of Markov Chain Monte Carlo (MCMC) methods. For this purpose we resort to large sets of data including measurements of the background expansion of the Universe, temperature fluctuations power spectra and those of gravitational potentials.

The manuscript is organised as follows. We lay down the theoretical framework in Sec.~\ref{Sec:theory}: the Kinetic model is introduced in Sec.~\ref{sec:model} and the explicit equations of motion for the background dynamics and linear scalar perturbations in the Newtonian gauge are presented in Sec.~\ref{sec:bckgrd} and Sec.~\ref{sec:prts}, respectively; in Sec.~\ref{Sec:Parspace} we discuss the parameter space of the model in order to guarantee its theoretical viability. 
In Sec.~\ref{Sec:Phenomenology} we focus on the cosmological properties of the Kinetic model, exploring the signatures left by the dark coupling on the background expansion in Sec.~\ref{sec:backevol}, and on the relevant cosmological observables in Sec.~\ref{sec:cosmoobs}. Finally, in Sec.~\ref{sec:obs} we present the observational constraints on the free cosmological and model parameters along with a model selection analysis.
Finally, we summarise our findings in Sec.~\ref{sec:conclusions}. Appendix~\ref{synchronous} provides the linear perturbation equations for the Kinetic model in the Synchronous gauge as well. 

%%%%%%%%%%%%%%%%%%%%%%%%%%%%%%%%%%%%%%%%%%%%%%%%%%%%%%%%%%%%%%%%%%%%%%

%%%%%%%%%%%%%%%%%%%%%%%%%%%%%%%%%%%%%%%%%%%%%%%%%%%%%%%%%%%%%%%%%%%%%%

\section{Theory}\label{Sec:theory}

In this Section we will present the theoretical formulation of the kinetically coupled dark energy model in consideration. We present the covariant formulation and the corresponding equations in Section~\ref{sec:model}, followed by the background evolution and the framework for linear scalar perturbations in Sections~\ref{sec:bckgrd} and \ref{sec:prts}, respectively. We then discuss the parameter space in Section \ref{Sec:Parspace}.

%-------------------------
\subsection{The Kinetic model}\label{sec:model}
%-------------------------

Let us start by considering a phenomenological theory minimally coupled to gravity in the Einstein frame, where the dark energy source is portrayed by a dynamical quintessence field, $\phi$, interacting with a dark matter component \textit{via} the action \cite{Barros:2019rdv},
\be\label{eq:action}
\mathcal{S} = \int {\rm d}^4 x \sqrt{-g} \left[ \frac{\text{M}_{\text{Pl}}^2}{2} R + X- V(\phi) + {f}(X)\tilde{\mathcal{L}}_c (\zeta,g_{\mu\nu} )+\mathcal{L}_\text{SM}(\psi_i,g_{\mu\nu}) \right] 
\mathcomma
\ee
where $g$ denotes the determinant of the metric tensor, $g_{\mu\nu}$, $R$ is the curvature scalar and ${\text{M}_{\text{Pl}}^2=(8\pi G)^{-1}}$ is the Planck mass in units of $c=1$, with $G$ being the Newtonian constant. The second and third terms in the action denote the scalar field Lagrangian, in which ${X=-g^{\mu\nu}\partial_{\mu}\phi\partial_{\nu}\phi}/2$ stands for the kinetic term of $\phi$ and $V(\phi)$ is the scalar self-interacting potential. In this work we extend the conventional quintessence formulation by taking a purely kinetic function, $f(X)$ multiplying the Lagrangian of cold dark matter, $\tilde{\mathcal{L}}_c$, which mediates a coupling of $\phi$ to the dark matter field $\zeta$.
Finally, $\mathcal{L}_\text{SM}(\psi_i,g_{\mu\nu})$ denotes a collective representation of Lagrangians of the uncoupled standard model fields, $\psi_i$.

Variation of the action in Eq.~\eqref{eq:action} with respect to the metric $g^{\mu\nu}$ yields the following field equations
\be\label{eq:field}
\text{M}_{\text{Pl}}^2 G_{\mu\nu}=T^{(\phi)}_{\mu\nu} + T^{(c)}_{\mu\nu} + T^{(b)}_{\mu\nu} + T^{(r)}_{\mu\nu} \mathcomma
\ee
with $G_{\mu\nu}$ being the Einstein tensor and $T^{(i)}_{\mu\nu}$ the energy momentum tensor for the $i$th species, defined as:	
\be
T^{(i)}_{\mu\nu} = -\frac{2}{\sqrt{-g}}\frac{\delta \left( \sqrt{-g}\mathcal{L}_i \right)}{\delta g^{\mu\nu}} \mathcomma
\ee
where $i=\phi,c,b,r$ and $c$ denotes the cold dark matter, $b$ the baryons and $r$ the radiation. 
Let us note that, for the previous definition to be valid for all the fluids present in theory, we define an effective dark matter Lagrangian as follows \cite{Barros:2019rdv,Koivisto:2005nr,Kase:2019veo}
\begin{equation}
{\mathcal{L}_c \equiv f(X)\tilde{\mathcal{L}}_c},
\end{equation}
incorporating the effect of the coupling.
We follow to consider that all the matter components in the theory can be modelled as perfect fluids, with energy density $\rho_i$, pressure $p_i$, and equation of state (EoS) parameter $w_i=p_i/\rho_i$. Therefore, the energy momentum tensor of each $i$th species becomes fully defined in terms of the fluid variables:
\be\label{eq:emtensor}
T^{(i)}_{\mu\nu}=\rho_i\left[\left(1+w_i\right)u^{(i)}_{\mu}u^{(i)}_{\nu} + w_i g_{\mu\nu}\right] \mathcomma
\ee
with $u^{(i)}_{\mu}$ being the 4-velocity vector associated with the $i$th species, under the individual constraint $g^{\mu \nu} {u^{(i)}_{\mu}u^{(i)}_{\nu}=-1}$. Regarding the EoS parameter, we have, $w_r=1/3$ for radiation, and $w_b=w_c=0$  for baryons and cold dark matter, respectively. In view of these considerations, the dark matter Lagrangian takes the particular form \cite{Koivisto:2005nr,Avelino:2018qgt},
\be\label{L_c_rho_c}
\mathcal{L}_c=-\rho_c\mathperiod
\ee
The scalar field admits a perfect fluid description as well \cite{Faraoni:2012hn}, provided that
\be 
u^{(\phi)}_{\mu}=-\frac{\partial_{\mu}\phi}{\sqrt{2X}} \mathcomma
\ee
and $X>0$, where the energy density and pressure associated to the quintessence field are given by:
\begin{eqnarray}
\rho_\phi&=& X+V\mathcomma\label{rho_phi}\\
p_\phi&=&X-V\mathperiod \label{p_phi}
\end{eqnarray}
The scalar field EoS parameter is 
\be \label{eq:wphi}
w_{\phi}=p_{\phi}/\rho_{\phi} \mathperiod
\ee

The equation of motion for the quintessence field, or simply the Klein-Gordon equation, is obtained through variation of the action in Eq.~\eqref{eq:action} with respect to $\phi$ and reads:
\be\label{eq:kleingordon}
\square\phi-V_{, \phi}=-Q \mathcomma
\ee
with ${V_{, \phi}=\dd V/\dd \phi}$. The term on the right-hand side of Eq.~\eqref{eq:kleingordon} includes the interaction in the dark sector in terms of $f(X)$ \cite{Barros:2019rdv}, and may be expressed as
\begin{eqnarray}
\label{coupling}
Q &=&-\mathcal{L}_c\left\{ \frac{{f}_{,X}}{{f}}\left[ \square\phi+\partial^{\mu}\phi\left( \frac{\nabla_{\mu}\mathcal{L}_c}{\mathcal{L}_c}  +\frac{{f}_{,X}}{{f}}\, \partial_{\alpha}\phi\nabla_{\mu}\partial^{\alpha}\phi \right)\right] - \frac{{f}_{,XX}}{{f}}\,\partial^{\mu}\phi\partial_{\alpha}\phi\left( \nabla_{\mu}\partial^{\alpha}\phi \right)\right\}\mathcomma
\end{eqnarray}
where  ${f_{,X}\equiv \dd f/\dd X}$ and ${f_{,XX}\equiv \dd^2f/\dd X^2}$.
The uncoupled case ($Q=0$) is naturally recovered when $f$ is a constant function. Let us note that Eq.~\eqref{eq:kleingordon} could likewise be found through the contracted Bianchi identities, yielding the following conservation relations,
\be\label{eq:consphim}
\covd_{\mu}T^{(c)}{}^{\mu}{}_{\nu}=-\covd_{\mu}T^{(\phi)}{}^{\mu}{}_{\nu}=Q\covd_{\nu}\phi \mathperiod
\ee
These equations illustrate clearly the energy transfer between the scalar field and DM when $f$ is not a constant, meaning that the dark components are not individually conserved. However, since radiation and baryons remain non-interacting, {\it i.e.},
\be\label{eq:consradbar}
\covd_{\mu}T^{(r)}{}^{\mu}{}_{\nu}=\covd_{\mu}T^{(b)}{}^{\mu}{}_{\nu}=0 \mathcomma
\ee
then, consistently, the overall energy momentum tensor of the theory is conserved, rendering the total action covariant.

In this work, we will focus on the case of a power-law interaction, motivated in \cite{Barros:2019rdv}, and parameterised by the function
\be\label{eq:coupling}
f(X)= \left( \text{M}_{\text{Pl}}^{-4}\, X \right)^{\alpha} \mathcomma
\ee
where $\alpha$ is a dimensionless constant.
Therefore Eq.~\eqref{coupling} becomes
\be\label{eq:Qterm}
Q = -\rho_c\frac{\alpha}{X}\left(\square\phi+\frac{\partial^{\mu}\phi\partial_{\nu}\phi\covd_{\mu}\partial^{\nu}\phi}{X}+\partial^{\mu}\phi\frac{\partial_{\mu}\rho_c}{\rho_c}\right)\mathperiod 
\ee
From Eq.~\eqref{eq:Qterm} it is straightforward to conclude that the parameter $\alpha$ governs the strength of the coupling within the dark sector. Additionally, we fully specify the model by considering the case of an exponential potential, that is,
\be\label{eq:pot}
V(\phi)=V_0 \expe^{-\lambda\phi/\text{M}_{\text{Pl}}} \mathcomma
\ee
where $V_0$ is the energy scale of the potential (a constant with dimensions of mass$^4$), and $\lambda$ is a dimensionless parameter depicting the steepness of the potential. The particular choices in Eqs.~\eqref{eq:coupling} and \eqref{eq:pot} are motivated by the possibility of having a scaling regime at early times, which is then followed by a period of accelerated expansion driven by $\phi$ \cite{Barros:2019rdv}. In terms of a dynamical systems analysis, the kinetic coupling is indeed responsible for the emergence of two novel critical points corresponding to scaling solutions. 
Finally, the role of the exponential potential is to drive the evolution of the system out of this scaling regime and towards the late time attractor. 

We conclude this section by remarking that the theory described by the action Eq.~\eqref{eq:action} is mathematically equivalent (namely it reproduces the same field equations and thus leads to equivalent cosmological dynamics) to that of the following scalar-tensor theory in the Einstein frame \cite{Brax:2016kin} 
\be\label{eq:scalartensor}
\mathcal{S} = \int {\rm d}^4 x \sqrt{-g} \left[ \frac{\text{M}_{\text{Pl}}^2}{2} R + X- V(\phi) +\mathcal{L}_i(\psi_i,g_{\mu\nu}) \right]+\mathcal{S}_c\left[\tilde{g}_{\mu\nu}(X),\zeta \right]\mathcomma
\ee
where $\tilde{g}_{\mu\nu}$ is the Jordan frame metric.
However, it is worth noting that their physical interpretation differs: while in the action \eqref{eq:action} the coupling is imposed directly through $f$, in the action \eqref{eq:scalartensor} the metric $\tilde{g}(X)$ defines {\it a posteriori} the coupling. In order for both theories to give rise to the same cosmological physics, the two metrics must be conformally related by the following Weyl scaling
\be
\tilde{g}_{\mu\nu}=f^2(X)g_{\mu\nu}\mathcomma
\ee
with conformal factor given by the square of $f$. Note that the square term automatically guarantees the signature of the metric to be preserved in both the Jordan and Einstein frames. These conformally coupled theories can also be written in terms of a non-minimal coupling to matter in the Einstein frame \cite{Pettorino:2008ez,Amendola:1999qq,Damour:1990tw}; nevertheless it is most common to assume a sole field dependence, {\it i.e.}, ${\tilde{g}_{\mu\nu}=\Omega(\phi)g_{\mu\nu}}$ \cite{Teixeira:2019tfi,Pettorino:2008ez,Barros:2018efl}. The mapping between the different formulations still applies as we have assumed in Eq.~\eqref{L_c_rho_c} that the cold dark matter on-shell Lagrangian can be described by its trace \cite{Avelino:2022eqm,Ferreira:2020fma,Avelino:2018rsb}, $T^c$, more generally,
\be\label{L_trace}
\mathcal{L}_c=T^c \equiv g^{\mu \nu} T^c_{\mu \nu} \mathperiod
\ee
If a different form for the nature of the cold dark matter Lagrangian had been adopted, departing from the perfect fluid description, then the relation in Eq.~\eqref{L_trace} might not hold, in which case the mapping between the theories would break down. Notice that the same power law coupling, ${f\propto X^{\alpha}}$, was considered in Ref.~\cite{Barros:2022kpo} to couple a quintessence field to Maxwell's electromagnetism. It was found that the model dynamics were mathematically equivalent to a disformally coupled theory. The reason for this stems from the fact that radiation is conformally invariant (since it has a vanishing energy-momentum trace ${T^r=0}$) thus one needs to consider a more general Weyl scaling such as to induce an interaction at the level of the field equations. This clearly shows that the correspondence between the theories strongly depends on the nature of the matter fields one wishes to couple the scalar source to.

%%%%%%%%%%%%%%%%%%%%%%%%%%%%%%%%%%%%%%%%%%%%%%%%%%%%%%%%%%%%%%%%%%%%%%

%%%%%%%%%%%%%%%%%%%%%%%%%%%%%%%%%%%%%%%%%%%%%%%%%%%%%%%%%%%%%%%%%%%%%%
%-------------------------
\subsection{Background equations}\label{sec:bckgrd}
%-------------------------

For what concerns the cosmological background dynamics, let us assume a flat Friedmann–Lema\^{i}tre–Robertson–Walker (FLRW) metric, expressed in terms of the conformal time $\tau$, as
\be\label{eq:backmetric}
\dd s^2 = a(\tau)^2\left( -\dd \tau^2 + \delta_{ij} \dd x^i \dd x^j \right) \mathcomma
\ee
where $a \equiv a(\tau)$ is the scale factor of the Universe.

The equations governing the background evolution can be derived from Eq.~\eqref{eq:field}, more precisely the modified Friedmann equation and the conservation relations, Eqs.~\eqref{eq:kleingordon}, \eqref{eq:consphim} and \eqref{eq:consradbar}, which become,
\begin{eqnarray}
3 \text{M}_{\text{Pl}}^2 \mathcal{H}^2&=&a^2(\rho_c+\rho_b+\rho_r+\rho_\phi)\label{eq:modfrid}\mathcomma\\
\phi'' + 2\mathcal{H}\phi'+a^2V_{,\phi} &=& a^2Q \label{eq:phiII} \mathcomma\\
\rho_c'+3\mathcal{H}\rho_c &=& -Q\phi' \mathcomma \label{eq:rhoII}\\
\rho'_b+3\mathcal{H}\rho_b &=& 0 \mathcomma \label{eq:rhoIII}\\
\rho'_r+4\mathcal{H}\rho_r &=& 0 \mathcomma \label{eq:rhoIV}
\end{eqnarray}
where a prime is used to refer to derivatives with respect to conformal time, $\mathcal{H}=a'/a$ is the Hubble rate in conformal time, and the coupling term in Eq.~\eqref{eq:Qterm} may now be written as:
\be\label{eq:kineticcoupling}
Q=2\alpha\rho_c\frac{3\mathcal{H}\phi'+a^2V_{, \phi}}{2\alpha a^2\rho_c + \left( 1+2\alpha \right)\phi'^2} \mathperiod
\ee
We can further define the energy density and pressure of the $\phi$ field at the background level, through Eqs.~\eqref{rho_phi} and \eqref{p_phi}, as
\begin{eqnarray}
\rho_{\phi} &=& \frac{\phi'^2}{2a^2}+V \mathcomma \label{eq:rhophi} \\
p_{\phi} &=& \frac{\phi'^2}{2a^2}-V \mathcomma \label{eq:pphi}
\end{eqnarray}
respectively. Therefore Eq.~\eqref{eq:phiII} can be written as:
\begin{equation}    \label{eq:contphi}
    \rho_\phi^\prime+3\mathcal{H}(1+w_{\phi})\rho_\phi=Q\phi^\prime.
\end{equation}
Equations \eqref{eq:rhoII} and \eqref{eq:contphi} imply that, when $Q\phi'>0$, energy is being transferred from the cold dark matter source to the scalar field, and, accordingly, the opposite holds when $Q\phi'<0$, and it is the $\phi$-field granting energy to cold dark matter. 
At the classical level the energy exchange in the dark sector may be interpreted as a mass variation for dark matter particles, since $m_c = a^3\rho_c$, assuming conservation of the number of particles, {\it i.e. $N_c=N_{c}(\tau_0)$}, with $\tau_0$ being the present conformal time. Integration of Eq.~\eqref{eq:rhoII} yields an expression for the total energy density of coupled dark matter,
\be\label{rho_c_an}
\rho_c = \rho_c(\tau_0)a^{-3}\exp \left(2\alpha\int_{\tau_0}^{\tau}Q\frac{\phi'}{\rho_c}{\rm d}\tau\right) \mathcomma
\ee
that can be expressed equivalently in terms of the mass of the dark matter particles:
\be\label{eq:DMmass}
m_{c}(\tau)=m_{c}(\tau_0)\exp \left(2\alpha\int_{\tau_0}^{\tau}Q\frac{\phi'}{\rho_c}{\rm d}\tau\right) \mathperiod
\ee

Finally let us note that the modified Friedmann equation, Eq.~\eqref{eq:modfrid}, can be cast to the form of the well-known Friedmann constraint:
\be\label{eq:friedmannback}
1=\Omega_\phi+\Omega_m+\Omega_r \mathcomma
\ee
where we have defined a collective matter density {$\rho_m = \rho_c + \rho_b$}, and the fractional density parameter of the $ith$ species $\Omega_i=\rho_ia^2/(3 \text{M}_{\text{Pl}}^2 \mathcal{H}^2)$. 
Eq.~\eqref{eq:friedmannback} can be rewritten in the form of a constraint on the present scalar field fractional density, $\Omega_{\phi}^0=1-\Omega_m^0-\Omega_r^0$, where ``0'' stands for quantities evaluated at the present time, $\Omega_i^0=\rho_i^0/(3 \text{M}_{\text{Pl}}^2 H_0^2)$, where $H_0$ is the Hubble parameter. For numerical purposes, $V_0$, implicitly entering the definition of $\Omega_\phi^0$, is used to perform a shooting method that yields the fiducial value of $\Omega_\phi^0$ fulfilling the constraint relation in Eq.~\eqref{eq:friedmannback}, while simultaneously avoiding degeneracies. As such, $V_0$ will no longer be considered a free parameter of the model, leaving $\{\lambda,\alpha \}$ as the model free parameters.

%-----------------------------
\subsection{Linear cosmological perturbations}\label{sec:prts}
%----------------------------

For the purpose of studying the background dynamics, we have assumed that the Universe is homogeneous and isotropic on large scales. However, we know that the global picture is far more complex and that, in particular, deviations to the homogeneous model are needed in order to explain phenomena such as the formation of structures in the Universe. For the purpose of this study, we consider small inhomogeneities of the geometry (encoded in the metric) and the matter fields, and investigate their synergy through the Einstein equations on linear scales. 

Let us consider the perturbed FLRW metric in the so called Newtonian gauge, corresponding to a line element written as follows \cite{Ma:1995ey}: 
\be\label{eq:newtonian}
 \dd s^2 = a^2(\tau)\left[ -\left( 1+2\Psi \right)\dd \tau^2 + \left( 1-2\Phi \right)\delta_{ij}\dd x^i \dd x^j \right] \mathcomma
\ee
where $\Psi(\vec{x},\tau)$ and $\Phi(\vec{x},\tau)$ are the Newtonian potentials. We  also consider linear perturbations around the relevant background fluid variables:
\begin{eqnarray}
\phi (\vec{x},\tau) = \phi_i(\tau) + \delta \phi(\vec{x},\tau) \mathcomma \quad
\rho_i (\vec{x},\tau) = \rho_i(\tau) + \delta \rho_i(\vec{x},\tau) \mathcomma \quad
p_i (\vec{x},\tau) = p_i(\tau) + \delta p_i(\vec{x},\tau) \mathperiod
\label{eq:pertquant}
\end{eqnarray}
In particular from Eqs.~\eqref{rho_phi} and \eqref{p_phi} we derive the perturbations for the energy density and pressure of the scalar field:
\begin{eqnarray}
\delta \rho_{\phi} &=& \frac{\phi'}{a^2}\delta\phi'-\frac{\phi'^2}{a^2}\Psi + V_{,\phi}\delta\phi \mathcomma \label{eq:deltaphi1}\\
\delta p_{\phi} &=& \frac{\phi'}{a^2}\delta\phi'-\frac{\phi'^2}{a^2}\Psi - V_{, \phi}\delta\phi \mathperiod
\end{eqnarray}

The perturbations of the energy-momentum tensor, Eq.~\eqref{eq:emtensor}, for each species and at first order, read
\be
\delta T^{(i)}{}^{\mu}_{\nu} = (\delta\rho_i+\delta p_i)u^{(i)}{}^{\mu}u^{(i)}{}_{\nu}+\delta p_i \delta^{\mu}_{\nu}+(\rho_i+p_i)\left( \delta u^{(i)}{}^{\mu}u^{(i)}{}_{\nu} + u^{(i)}{}^{\mu} \delta u^{(i)}{}_{\nu} \right) \mathcomma
\ee
where $\delta u^{(i)}{}_{\mu}$ is the perturbation on the four velocity vector of the $i$th-species, {\it i.e.} ${u^{(i)}{}_{\mu}=a(-1,v^{(i)}{}_j)}$, with $v_j$ being the peculiar velocity. In this study we will assume that there is no anisotropic stress associated with the fluids under consideration.

Then we compute the linearised Einstein equations where we include the modifications introduced by the coupling function. These are expressed in terms of independent Fourier modes that characterise the evolution of the perturbations for different scales:
\begin{eqnarray}
k^2 \Phi+3\mathcal{H}\left(\Phi'+\mathcal{H}\Psi\right) &=& -4\pi G a^2 \sum_i \delta \rho_i \label{eq:lEE1} \mathcomma \\
k^2\left(\Phi'+\mathcal{H}\Psi\right) &=& 4\pi G a^2 \sum_i \rho_i (1+w_i) \theta_i \mathcomma \label{eq:lEE2}  \\
\Phi''+\mathcal{H}\left(\Psi'+2\Phi'\right)+\Psi\left(\mathcal{H}^2+2\mathcal{H}'\right)+\frac{k^2}{3}\left( \Phi-\Psi \right)&=&4\pi G a^2\sum_i\delta p_i \mathcomma \label{eq:lEE3} \\
\Phi &=& \Psi \mathperiod \label{eq:lEE4}
\end{eqnarray}
The first equation, corresponding to the time-time component, provides the energy density constraint.
Equation \eqref{eq:lEE2}, computed from the time-space components of the perturbed Einstein equations, gives the momentum constraint, where we have adopted the definition of the velocity divergence ${\theta_i = \covd \cdot v^{(i)}}$. The trace of the spatial components yields Eq.~\eqref{eq:lEE3} and, finally, Eq.~\eqref{eq:lEE4} corresponds to the shear propagation for vanishing anisotropic stress. This relation is expected due to the lack of a non-minimal coupling in action (\ref{eq:action}).

The equations governing the evolution of each fluid's perturbations can be found through the conservation relations, Eqs.~\eqref{eq:consphim} and \eqref{eq:consradbar}, perturbed at first order. For the non-interacting species, \textit{i.e.} baryons and radiation, these are respectively
\begin{eqnarray}
\delta'_i+3\mathcal{H}\left( \frac{\delta p_i}{\delta \rho_i}-w_i \right)\delta_i + (1+w_i)\left( \theta_i -3\Phi' \right) &=& 0 \mathcomma \label{eq:nonintdelta} \\
\theta_i' +\left[ \mathcal{H}(1-3w_i)+\frac{w_i'}{1+w_i} \right]\theta_i-k^2\left( \Psi + \frac{\delta p_i}{\delta \rho_i}\frac{\delta_i}{1+w_i}\right) &=& 0 \mathcomma
\end{eqnarray}
where we have defined the dimensionless density contrast as ${\delta_i=\delta \rho_i/\rho_i}$. The dynamics for the  coupled cold dark matter is given by
\be\label{eq:cdmdelta}
\delta_c' + \theta_c -3\Phi' = \frac{Q}{\rho_c} \left( \phi'\delta_c - \delta\phi' \right) - \frac{\phi'}{\rho_c}\delta Q \mathcomma
\ee
and the corresponding velocity divergence evolves according to
\be\label{eq:cdmtheta}
\theta'_c + \mathcal{H}\theta_c -k^2\Psi= \frac{Q}{\rho_c}\left( \phi'\theta_c - k^2\delta\phi \right) \mathcomma
\ee
with the perturbed coupling term, obtained from Eq.~\eqref{eq:Qterm}, being defined as 
\begin{eqnarray}
\delta Q &=& \frac{2\alpha\rho_c}{2\alpha a^2 \rho_c + (1+2\alpha)\phi'^2}\left\{ -3\Phi'\phi' -\phi'\theta_c + \left[ 3\mathcal{H}\phi' + a^2 (V_{,\phi}-Q) \right]\delta_c + \left( 2k^2 + a^2 V_{,\phi\phi} \right)\delta\phi \right. \nonumber \\
&&\left. \hspace{3.8cm} -\left[ 3\mathcal{H}\phi' + 2a^2 (V_{,\phi}-Q) \right] \frac{\delta\phi'}{\phi'} +2a^2\Psi\left(Q-V_{,\phi}\right)\right\} \mathcomma\label{eq:deltaQ}
\end{eqnarray}
with $V_{,\phi\phi} = \dd^2 V / \dd \phi^2$.
One exceptional feature of the Kinetic model can be readily identified at the level of the perturbed coupling parameter, Eq.~\eqref{eq:deltaQ}: it includes an explicit dependence on $\theta_c$. This is not usual in other coupled dark energy models explored so far, such as in Refs.  \cite{vandeBruck:2016hpz,vandeBruck:2020fjo}, and it arises due to the $X$-dependence of the coupling, in particular in relation to the term containing ${\covd_{\mu}\mathcal{L}_c}$ in Eq.~\eqref{eq:Qterm}. 

The evolution of the $\phi$-field perturbation is given by the linearisation of Eq.~\eqref{eq:kleingordon}:
\be\label{eq:perturbedKG}
\delta\phi'' + 2\mathcal{H}\delta\phi' + \left( a^2V_{,\phi\phi} +k^2 \right)\delta\phi-\left(\Psi'+3\Phi'\right)\phi'+2a^2\Psi V_{,\phi} = a^2 \delta Q+2a^2Q\Psi \mathperiod
\ee

For completeness we also provide the corresponding set of equations in the synchronous gauge, see Appendix \ref{synchronous}. 

In Sections \ref{Sec:Phenomenology} and \ref{sec:obs} we will evolve the full dynamics of the gravitational potentials, scalar and matter fields. However in order to have a glimpse at the phenomenology of the Kinetic model we resort to the  so-called quasi-static approximation (QSA)  on sub-horizon scales \cite{Boisseau:2000pr,Tsujikawa:2007gd,DeFelice:2011hq}. Under this approximation we find:
\begin{eqnarray}
&&k^2 \Psi \approx -
\frac{3\mathcal{H}^2}{2}\big(\Omega_c\delta_c+ \Omega_b\delta_b\big)\mathcomma\label{Eq:Poisson}\\
&&k^2\delta\phi\approx a^2\Delta \bigg\{ \delta_c \Big[Q  \phi '^2+\rho _c\big(\phi''-\mathcal{H}\phi'\big)\Big]-\rho_c\phi'\delta_c'\bigg\} \mathcomma
\end{eqnarray}
where  we have defined the following time and scale dependent functions:
\begin{eqnarray}
\Delta&=&-\frac{2\alpha\mathcal{M}^2}{\phi'^2V_{\phi\phi}\left(1+\frac{a^2}{k^2}\mathcal{M}^2\right)}\,,\\
\mathcal{M}^2&=&\frac{\phi'^2V_{\phi\phi}}{\phi'^2-2\alpha a^2\rho_c}\mathperiod
\end{eqnarray}
It follows that the equations for cold dark matter and baryonic fluids read
\begin{eqnarray}
&&\delta_c''+\mathcal{H}\big(1+\beta\big)\delta_c'-
\frac{3\mathcal{H}^2}{2G}\big(G_{cc}\Omega_c\delta_c+G_{cb}\Omega_b\delta_b\big)\approx 0    \mathcomma\label{Eq:deltacperturb}\\
&&\delta_b^{\prime\prime}+\mathcal{H}\delta'_b-
\frac{3\mathcal{H}^2}{2}\big(\Omega_c\delta_c+ \Omega_b\delta_b\big)\approx 0\mathcomma
\end{eqnarray}
with 
\begin{eqnarray}
\beta &=& \frac{G_{cb}}{G\mathcal{H}\rho_c}\bigg\{\big(1-a^2\rho_c\Delta\big)\Big[ 8\alpha\mathcal{H}\rho_c - Q\phi'\big(1+2\alpha\big)\Big]-Q\phi'\bigg\}\mathcomma \\
G_{cc} &=& G+\frac{2G}{3\mathcal{H}\rho_c\Omega_c\phi'^2}\bigg\{\phi'^2\Big[\phi'Q'+Q\big(\phi''+4\mathcal{H}\phi'\big)\Big]+\Big[Q\phi'^2+\rho_c\big(\phi''-\mathcal{H}\phi'\big)\Big]\\
&&\times\Big[ \phi'^2Qa^2\Delta + 2\alpha\big(\phi''+4\mathcal{H}\phi'\big)\big(1-a^2\rho_c\Delta\big) \Big]\Bigg\}\mathcomma\\
G_{cb} &=&\frac{G}{1+2\alpha\big( 1-a^2\rho_c\Delta \big)}\mathperiod
\end{eqnarray}
The cold dark matter perturbations are then modified by two effects emerging from the coupling: a modified friction term, quantified by $\beta$, which inevitably influences the growth rate of $\delta_c$; and a modified effective gravitational potential encoded in 
\be
\nabla^2\Psi^{\rm eff} = 4\pi  \big( G_{cc}\rho_c\delta_c + G_{cb}\rho_b\delta_b\big)\mathperiod
\ee
The latter includes two effective gravitational couplings, $G_{cc}$ and  $G_{cb}$, defined in analogy to \cite{Kase:2019veo}. We also find that $G_{cb}$ is always an attractive contribution. These modifications clearly show the emergence of a fifth force which is a standard signature of coupled scalar field models. We expect  that even a relative small value for the coupling parameter $\alpha$ can lead to a significant effect on the cosmological observables given the evolution equation for the cold dark matter perturbations, which impacts the baryons dynamics and the gravitational potentials.  Among others we foresee a modification in the lensing angular power spectrum due to a modified  lensing potential ($\phi_{lens}=(\Phi+\Psi)/2=\Psi$). We will explore these signatures in more detail in Section \ref{Sec:Phenomenology}.  In the absence of the coupling, \textit{i.e.} $\alpha=0$, we recover $\beta=0$ and $G_{cc}=G_{cb}=G$, corresponding to the standard case of quintessence.

%%%%%%%%%%%%%%%%%%%%%%%%%%%%%%%%%%%%%%%%%%%%%%%%%%%%%%%%%%%%%%%%%%%%%%

%-------------------------
\subsection{The parameter space}\label{Sec:Parspace}
%-------------------------

In order for a model to be theoretically viable, there are specific stability requirements that need to be satisfied. We will present and examine them below for the Kinetic model.
Let us stress that the identification of a physically motivated parameter space plays an important role when testing particular gravity models with cosmological data \cite{Raveri:2014cka,Frusciante:2015maa,Salvatelli:2016mgy,Frusciante:2018vht,Frusciante:2019puu,Frusciante:2019xia,Albuquerque:2021grl}.

According to the results in Ref.~\cite{Barros:2019rdv} the solutions with $\lambda^2<2$ guarantee that the future dark energy attractor is a stable fixed point of the system and describes an accelerated expanding Universe. Albeit necessary {\it at the attractor}, this condition can be somewhat relaxed, while still generating an accelerating behaviour at transient times. By allowing the attractor to lie outside, but close to the accelerated region, with say ${\lambda^2=2+\epsilon}$, the solution may still feature an accelerated expanding scenario at present time, that is, with ${w_{\phi}(a_0)< -1/3}$. We further discuss this point in Section~\ref{sec:backevol} and we show some examples in the right panel of Fig.~\ref{fig:fig3}. Let us note that under such condition, instead of accelerating forever, there should be a turning point in the future when the expansion changes from accelerated to decelerated, {\it i.e.} ${w_{\phi}(a_{x})=-1/3}$ at the crossover $a_{x}$, and ${w_{\phi}(a)>-1/3}$ thereafter, for $a<a_{x}$, as the attractor is approached. From the critical points analysis conducted in Ref.~\cite{Barros:2019rdv} we know that at the attractor ${w^{\star}_{\rm eff}=\lambda^2/3-1}$, which using the following general identity
\be
\frac{1}{H}\frac{\dd H}{\dd\ln a}=-\frac{3}{2}\left(1+w_{\rm eff}\right)\mathcomma
\ee
we find for the Hubble rate at the attractor:  %
\be
\frac{\dd H^{\star}}{\dd \ln a} = -\frac{1}{2}H^{\star}\lambda^2\mathcomma
\ee
from which the accelerated condition is derived \cite{Barros:2019rdv}. Here we define a star superscript denoting quantities evaluated at the attractor, and $H=\mathcal{H}/a$ is the Hubble function in cosmic time, $t$.
The relation above corresponds to a cosmological expanding behaviour described as,
\be
H^{\star}=H_0a^{-\lambda^2/2}\quad\text{and therefore}\quad a^{\star}= \left( 1+H_0\frac{\lambda^2}{2}t \right)^{2/\lambda^2}\mathperiod\label{eq:at_attractor}
\ee
Indeed, the explicit time dependence of the scale factor in Eq~\eqref{eq:at_attractor}, $a\propto t^{2/\lambda^2}$, reveals that $\lambda^2=2$ is an inflection point of $a(t)$, \textit{i.e.} $\ddot{a}=0$ (with dots referring to derivatives with respect to cosmic time), laying out the fine limit between an accelerated or decelerated setting. Following the above discussion we will then consider $\lambda>0$. Additionally, according to the power-law role of $\alpha$ in Eq.~\eqref{eq:coupling}, we choose to consider cases with $\alpha\geqslant 0$ only \cite{Barros:2019rdv}.  
 
Furthermore, we take into account theoretical stability conditions to guarantee the absence of ghost and gradient instabilities in the scalar sector~\cite{Sbisa:2014pzo,Kase:2019veo}. The first demands for positive kinetic terms of the scalar field and cold dark matter perturbations ($q_s>0$ and $q_c>0$, respectively), and the second for their positive speeds of propagation ($c_s^2\geqslant0$ and $c_c^2\geqslant0$). % In detail, for our case these read~\cite{Kase:2019veo}: 
It is possible to show that a very general way to write an action with an extra scalar field and one matter component up to second order in perturbations, is the following~\cite{Kase:2019veo}:
\begin{equation}
    \mathcal{S}^{(2)}=\int{}dt dk^3a^3\left[\dot{\vec{\chi}}^t{\bf K} \dot{\vec{\chi}}-\frac{k^2}{a^2}\vec{\chi}^t{\bf G} \vec{\chi}-\vec{\chi}^t{\bf M} \vec{\chi}-\frac{k}{a}\vec{\chi}^t{\bf B} \dot{\vec{\chi}} \right]
\end{equation}
with $\vec{\chi}^t=(\delta\phi,\delta\rho_c/k)$ being $\delta \phi$ and $\delta \rho_c$ the perturbations  of the scalar field and cold dark matter component respectively, the $2\times 2$ matrices are defined in terms of background quantities and their general forms can be found in ref.~\cite{Kase:2019veo}. For the action (\ref{eq:action}), with $f(X)$ and $V(\phi)$ defined  in Eqs. (\ref{eq:coupling}) and (\ref{eq:pot}) respectively we have~\cite{Kase:2019veo}:
\begin{eqnarray}\label{eq:scalarghost}
q_s &=& K_{11}= 2 \text{M}_{\text{Pl}}^2 \left[1-\frac{\alpha(2\alpha-1)}{X}\rho_c\right]
\mathcomma \label{q_s}\\
q_c&=& K_{22}=\left(\text{M}_{\text{Pl}}^{-4} X \right)^\alpha\mathcomma \label{q_c}\\
c_s^2&=& \frac{G_{11}}{K_{11}}+\frac{B_{12}^2}{K_{11}K_{22}}= \frac{4\text{M}_{\text{Pl}}^2}{q_s}-1\mathcomma \label{c_s_s} \\
c_c^2&=&\frac{G_{22}}{K_{22}}=0\mathperiod \label{c_s_c}
\end{eqnarray}
%
%------------
For dark matter the conditions are trivial.
The stability  conditions for the scalar field are more involved and need to be verified throughout the entire expansion history. We find that both conditions are verified as long as:
\be\label{scalar_condition}
-1<\alpha (2\alpha -1)\frac{\rho_c}{X}<1\mathcomma
\ee
where the first inequality accounts for the no-ghost condition and the second one for the positive (square) speed of propagation. This constraint then selects the viable range for the parameter $\alpha$. The initial condition  $\phi'_i$  plays a role in securing the stability toward  the cosmological evolution.

Let us now discuss the initial conditions (ICs) for the scalar field, $\phi_i$ and its first derivative, $\phi^\prime_i$ which must be specified, deep in the radiation dominated epoch, namely around redshift $z_i\approx 10^{14}$ in order to solve the system of equations (\ref{eq:phiII})-(\ref{eq:rhoIV}).

From the numerical study we concluded that for non-trivial ICs, the system rapidly enters in the scaling regime. According to this feature we found that the choice of values for $\phi_i$ and $\phi^\prime_i$ has a negligible impact on the cosmological evolution\footnote{We have numerically verified that the phenomenology of the cosmological observables, as discussed in Section \ref{Sec:Phenomenology}, is not affected by the choice of ICs and neither are the cosmological constraints.}. Hence, without loss of generality, we set $\phi(z_i)=10^{-2}\, \text{M}_{\rm Pl}$ .

The IC for $\phi'$ is chosen such as to avoid instabilities according to Eq.~\eqref{scalar_condition}.
Moreover, it should be noted that when $\phi^\prime_i$ is chosen to be positive, the condition $\lambda>0$ must hold for the accelerating attractor solution to exist \cite{Barros:2019rdv}.

Finally, we recall that $V_0$ is not  considered an extra parameter  of the model as discussed in Section \ref{sec:bckgrd}.

%-------------------------
\section{Phenomenology of the Kinetic model} \label{Sec:Phenomenology}
%-------------------------

In this Section we shall explore the signatures left by the Kinetic model on the background expansion and on some cosmological observables such as the cosmic microwave background (CMB), the lensing potential auto-correlation and the matter power spectra in Sections~\ref{sec:bckgrd} and \ref{sec:prts}, respectively. We use our own modification of the public version of the Einstein Boltzmann solver \texttt{CLASS} \cite{lesgourgues2011cosmic,Blas_2011,lesgourgues2011cosmic2}.
 
\begin{figure}
      \subfloat{\includegraphics[height=0.4\linewidth]{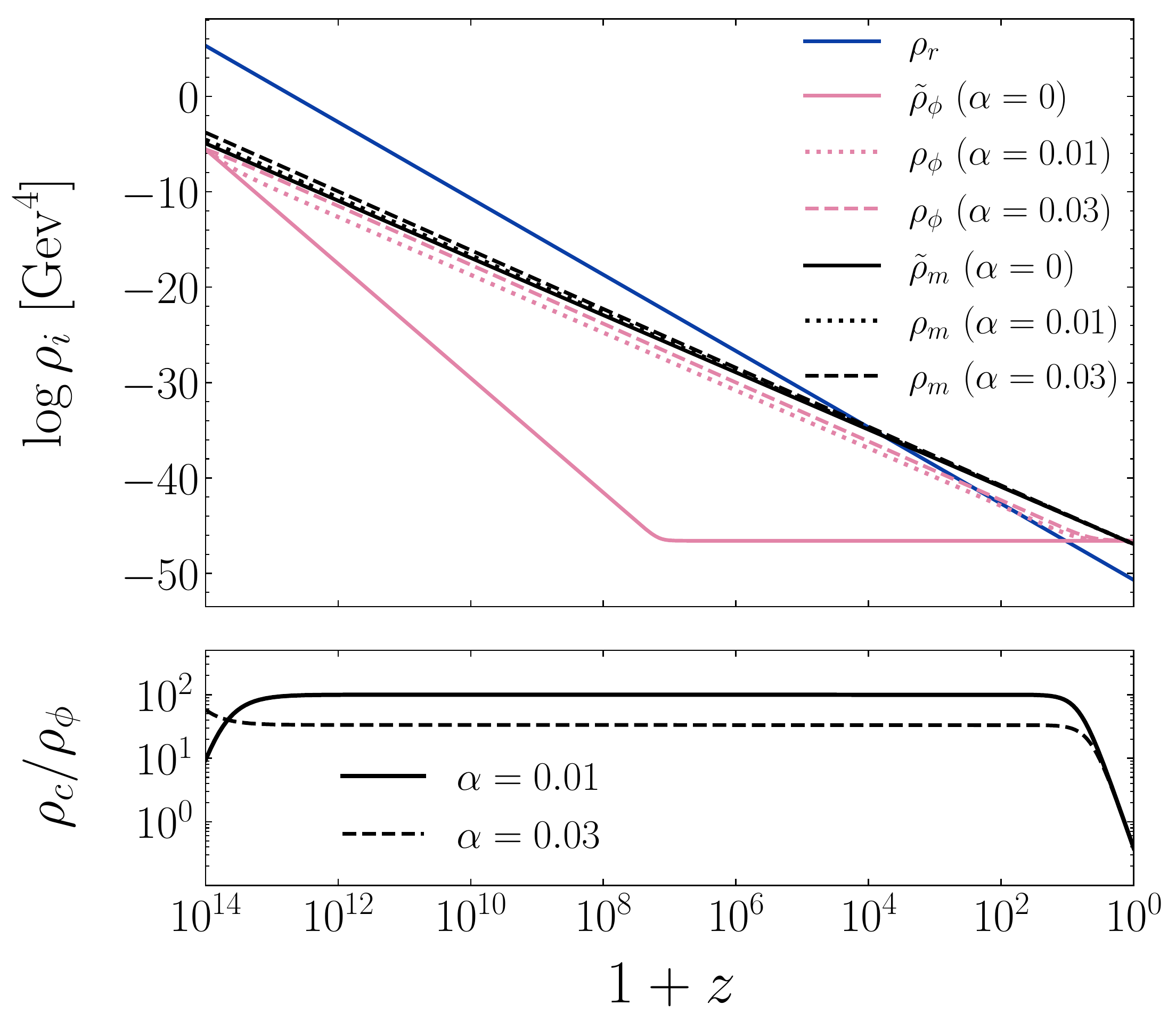}}
      \qquad
      \subfloat{\includegraphics[height=0.4\linewidth]{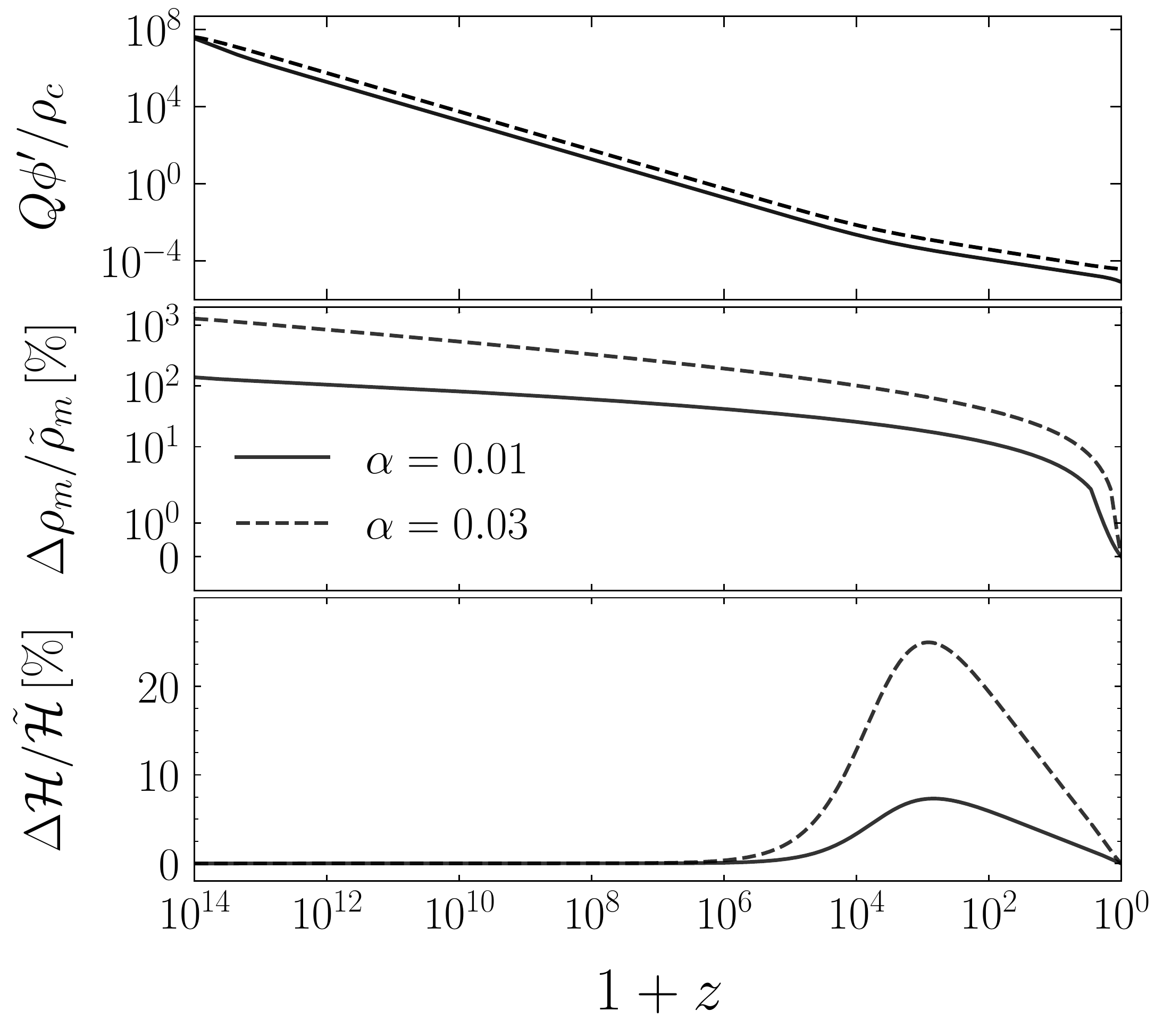}}
  \caption{\label{fig:fig1} \underline{Left upper panel}: Evolution of the energy densities  $\rho_i$ with redshift, $1+z$, of the scalar field (pink), matter (black) and radiation (blue)  for the uncoupled case (solid line), ${\alpha =0.01}$ (dotted line) and ${\alpha = 0.03}$ (dashed line).  \underline{Left lower panel}: Ratio of the energy densities of cold dark matter and dark energy, for ${\alpha = 0.01}$ (solid line) and ${\alpha = 0.03}$ (dashed line). \underline{Right panel}: Differences relative to the uncoupled case, ${\alpha =0}$, for ${\alpha = 0.01}$ (solid line) and ${\alpha = 0.03}$ (dashed line), on the quantities (from top to bottom): the coupling strength parameter; the fractional deviation of the energy density of matter, {\it i.e.} ${\Delta \rho_m/\tilde{\rho}_m = \rho_m/\tilde{\rho}_m - 1}$, where a tilde denotes variables in the uncoupled scenario, such that ${\tilde{\rho}_m = \rho^0_m a^{-3}}$; and the fractional deviation of the Hubble rate of expansion.}
\end{figure}

\subsection{Background evolution}\label{sec:backevol}
%-------------------------

We start by reviewing the background evolution in the kinetic coupled dark sector scenario. A similar study has been previously presented in Ref.~\cite{Barros:2019rdv} by means of a dynamical systems analysis, with a particular focus on the late time dynamics (\textit{i.e.} cosmological redshift $z\lesssim 40$), neglecting the radiation and baryonic contribution. In this work we shall examine the cosmological evolution starting from the early stages, deep into the radiation dominated epoch ($z_i\approx 10^{14}$) up to present time ($z = 0$). For the numerical investigation in this work we fix the following cosmological parameters to be \cite{Planck:2018vyg}: ${H_0=67.56}$ km/s/Mpc, ${\Omega_b h^2=0.022}$ and ${\Omega_c h^2=0.12}$, with ${h \equiv H_0/100}$. We also select some exemplifying values for the parameter $\alpha$ controlling the coupling, namely ${\alpha=0.01}$ and ${\alpha=0.03}$, and we fix the slope of the potential as ${\lambda=0.2}$, with the aim of singling out the main phenomenology associated to the coupling function. Moreover, for comparison purposes, we also include the case with $\alpha=0$, which corresponds to an uncoupled scenario. It should be noted that the choice for the values of the parameters associated with the scalar field are purely illustrative, but nevertheless still satisfy the requirements discussed in Section \ref{Sec:Parspace}. They are chosen in such a way that the overall effect of the coupling can be grasped, and therefore are not necessarily realistic. This will be assessed in Section \ref{sec:obs}, in which case these parameters are left to vary when performing a parameter estimation according to cosmological data.

In the left panel of Fig.~\ref{fig:fig1} we show the evolution with redshift, $1+z$, of the energy densities for each species, $\rho_i$. We notice that the introduction of the coupling results in the emergence of an early scaling regime, in direct contrast with the uncoupled case, for which this behaviour can never be achieved. The onset of this scaling behaviour takes place during the radiation dominated epoch, with energy density of the scalar field proportional to the dark matter one, approximately according to the relation ${\rho_c/\rho_{\phi}=1/\alpha}$, as shown in the left lower panel of Fig.~\ref{fig:fig1}. Eventually the field will exit this scaling regime and head towards the future attractor solution, in which case its energy density will remain forever diluting as ${\rho_{\phi}\propto a^{-\lambda^2}}$.

%--------------
In the upper right panel of Fig.~\ref{fig:fig1} we show the evolution of the coupling strength, expressed as $Q\phi'/\rho_c$, as a function of the redshift. The sign of this quantity is relevant to assess the direction of the energy flow between cold dark matter and the scalar field.

\begin{figure}
      \subfloat{\includegraphics[height=0.31\linewidth]{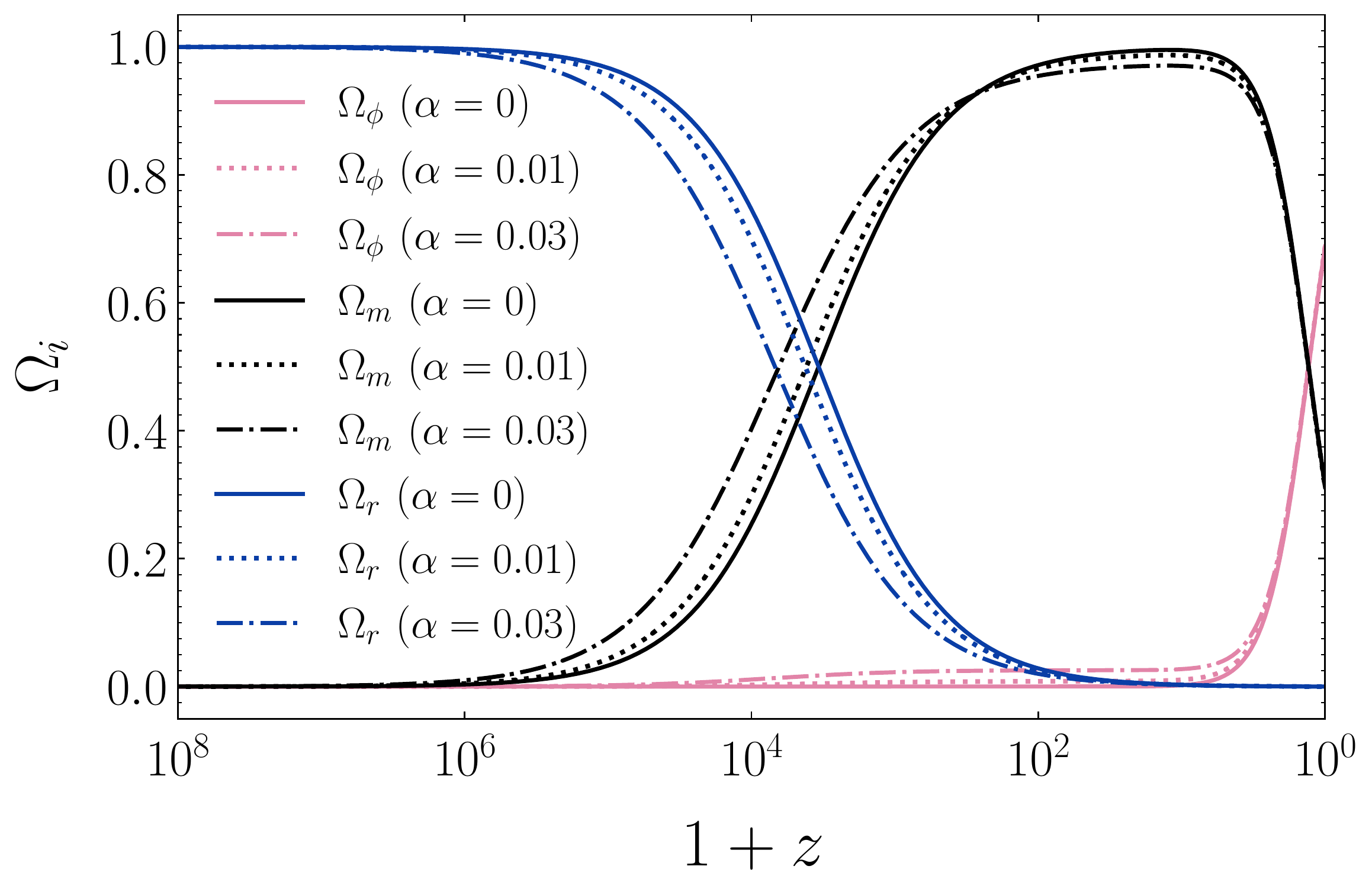}}
      \qquad
      \subfloat{\includegraphics[height=0.31\linewidth]{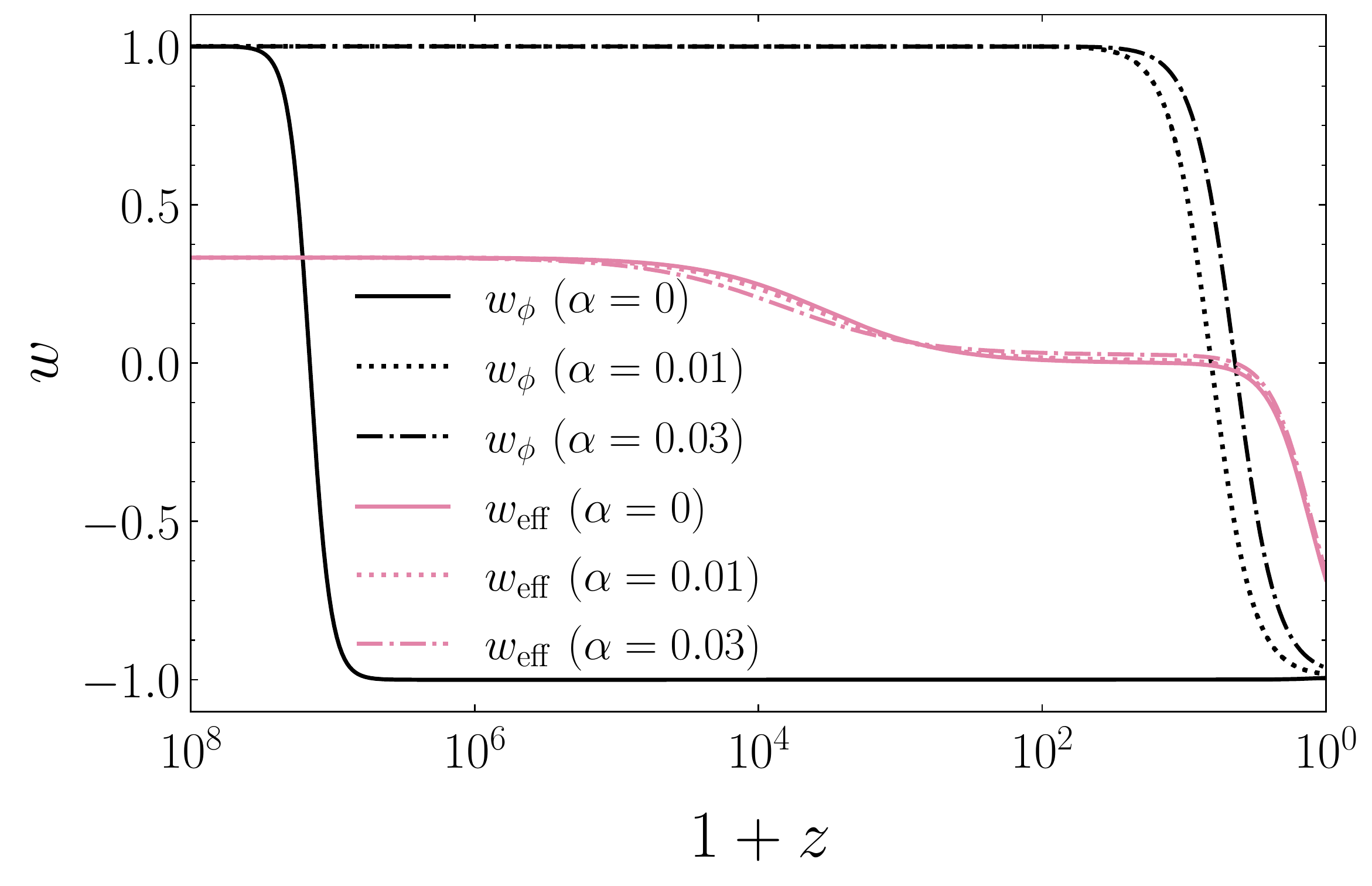}}
  \caption{\label{fig:fig2} \underline{Left panel}: Evolution of the relative energy densities  $\Omega_i$ with redshift, $1+z$, of the scalar field (pink), matter (black) and radiation (blue). \underline{Right panel}: Equation of state parameters, $w_{\rm eff}$ (pink) and ${w_{\phi}}$ (black), along redshift. In accordance with Fig.~\ref{fig:fig1}, we present the uncoupled case (solid line), ${\alpha =0.01}$ (dotted line) and ${\alpha = 0.03}$ (dot-dashed line).}
\end{figure}

We can notice that the interaction term is positive at all redshifts, establishing the direction of the energy transfer from the dark matter fluid to the scalar field. This is consistent with the fact that the dynamics of the scalar field follows the relation ${\phi' > \lambda V/\left( 3 \text{M}_{\text{Pl}}^2 \mathcal{H} \right) \Leftrightarrow Q>0}$ (see Eq.~\eqref{eq:kineticcoupling}). 
Let us note that because we fixed the present day values of the fluid densities this results in a larger value for the cold dark matter energy density at early times because it is the CDM component granting energy to the scalar field at later times, with this feature being more prominent for higher values of $\alpha$. This effect is compensated as the matter energy density decreases throughout time, while additional energy is being transferred for the scalar field, when compared with the uncoupled case. We illustrate this behaviour in the middle right panel of Fig.~\ref{fig:fig1}, where we report on the deviations from the uncoupled case, denoted by a tilde. As consequence there is a shift of the matter-radiation equality towards earlier times for increasing values of $\alpha$, as shown in the left panel of Fig.~\ref{fig:fig2}. From the same Figure, we can notice that because the $\phi$ field is acquiring energy at a rate that is proportional to its energy density (see Eq.~\eqref{eq:kineticcoupling}), then the matter-dark energy equality is achieved earlier.

Additionally, in the lower right panel of Fig.~\ref{fig:fig1}, we show the deviations in the Hubble rate for the Kinetic model when compared with the uncoupled case, {\it i.e.} ${\Delta \mathcal{H}/\tilde{\mathcal{H}} = \mathcal{H}/\tilde{\mathcal{H}}-1}$. No significant deviations on $\mathcal{H}$ are observed during the radiation dominated epoch, since any interactions between the dark and radiation sectors have been excluded. However, when the matter contribution becomes non-negligible, around ${z\approx 10^6}$, the Kinetic models show an enhanced value of $\mathcal{H}$ with respect to the uncoupled case, with this effect being larger for the higher values of $\alpha$.

Finally, it is also worth analysing the evolution of two fluid-related quantities: the equation of state parameters for the scalar field, $w_{\phi}$, and for the total effective budget,
$w_{\rm eff}$. These characterise the nature of the dark energy fluid description and the overall effective dominating fluid contribution in the Universe, and are defined according to Eq.~\eqref{eq:wphi} and
\be\label{w_eff}
w_{\rm eff}=\frac{\sum_i p_i}{\sum_i\rho_i}\mathcomma
\ee
respectively. Their evolution with redshift is depicted in the right panel of Fig.~\ref{fig:fig2}. We observe that during the scaling regime the field behaves as a stiff fluid, with ${w_{\phi}=1}$, since ${V\ll \phi'^2}$, and in agreement with the findings of Ref.~\cite{Barros:2019rdv}. As the field exits the scaling regime, the Universe approaches the attractor scenario, for which ${w_{\phi}=-1+\lambda^2/3}$. During radiation domination, the effective equation of state remains at a \textit{plateau} with ${w_{\rm eff}\approx w_r = 1/3}$. At matter domination, and during the scaling regime, when radiation may be neglected and under the limit ${V\ll\phi'^2}$, the equation of state follows
\be\label{eq:approx}
w_{\rm eff}\approx \frac{\alpha}{1+\alpha\left( 1+\frac{\rho_b}{\rho_{\phi}} \right)} \mathperiod
\ee
Note that, in Ref.~\cite{Barros:2019rdv}, a similar approximation was presented, though stated as ${w_{\rm eff}\approx \alpha/(1+\alpha)}$. That is because the contribution of radiation and baryons was not taken into account in that study, which focused mainly on the late time dynamics, for which it still stands as a good approximation. By neglecting the baryonic contribution we may resort to the dynamical system analysis employed in Ref.~\cite{Barros:2019rdv} to find the behaviour of the Hubble rate and coupled DM at matter domination during the scaling:
\be
\rho_c\propto H^2\propto a^{-3\frac{1+2\alpha}{1+\alpha}}\mathcomma
\ee
which we numerically verified to be a good approximation. We remark that the transition towards an accelerating state occurs later for increasingly larger values of $\alpha$, owing to the fact that, for a stronger interaction, the field remains frozen in the scaling regime for longer, with ${w_{\phi}=1}$. As a direct outcome, when the accelerating stage finally starts (that is, when $w_{\rm eff}<-1/3$), it will take place at a slower rate. This behaviour is illustrated in the right panel of Fig.~\ref{fig:fig2}. Alternatively, this trend could be intuitively understood by inspection of the deceleration parameter ${q=(1+3w_{\rm eff})/2}$, that scales linearly with the total equation of state parameter of the Universe.

\begin{figure}[t!]
      \subfloat{\includegraphics[height=0.305\linewidth]{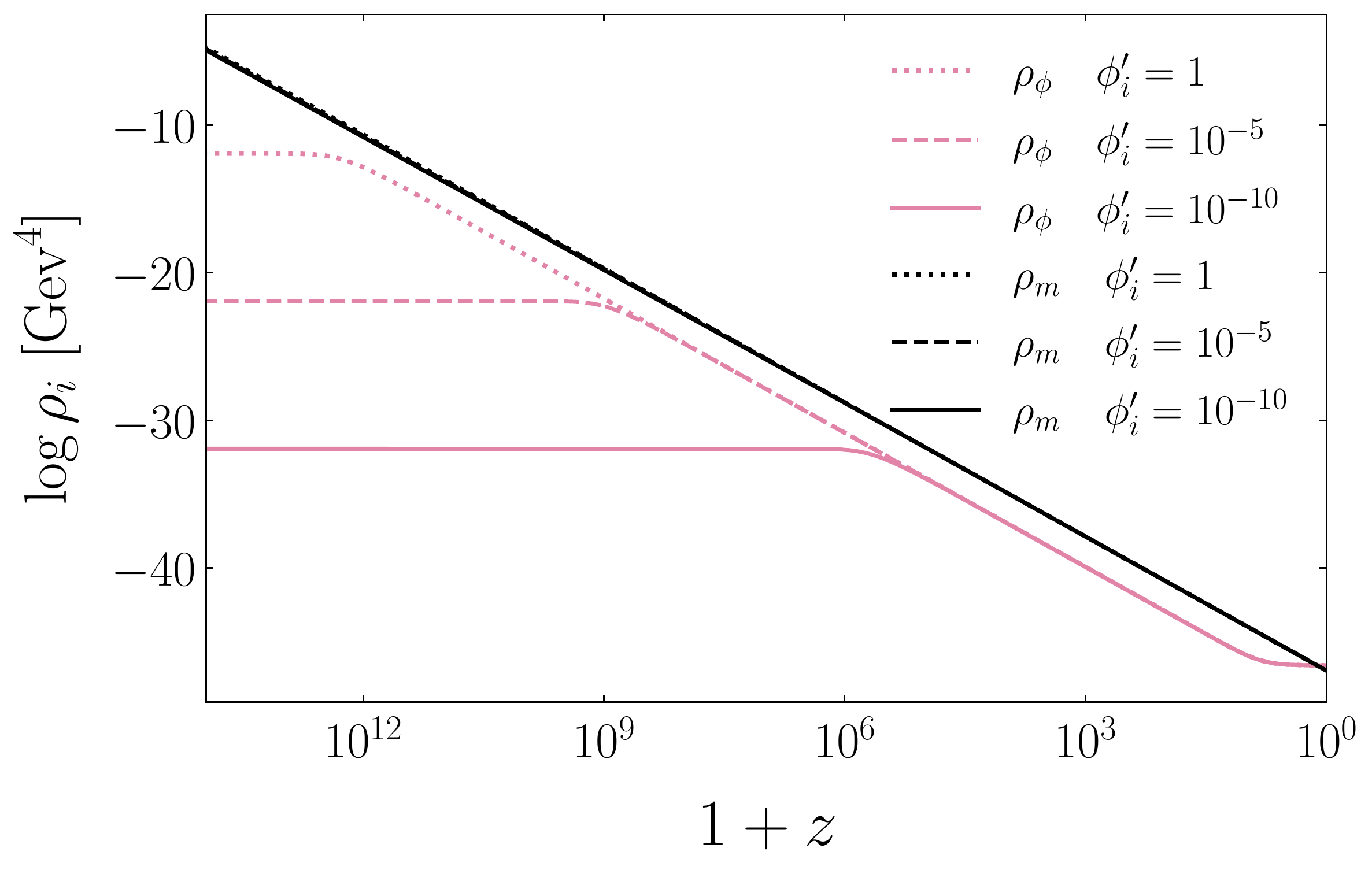}}
      \qquad
      \subfloat{\includegraphics[height=0.305\linewidth]{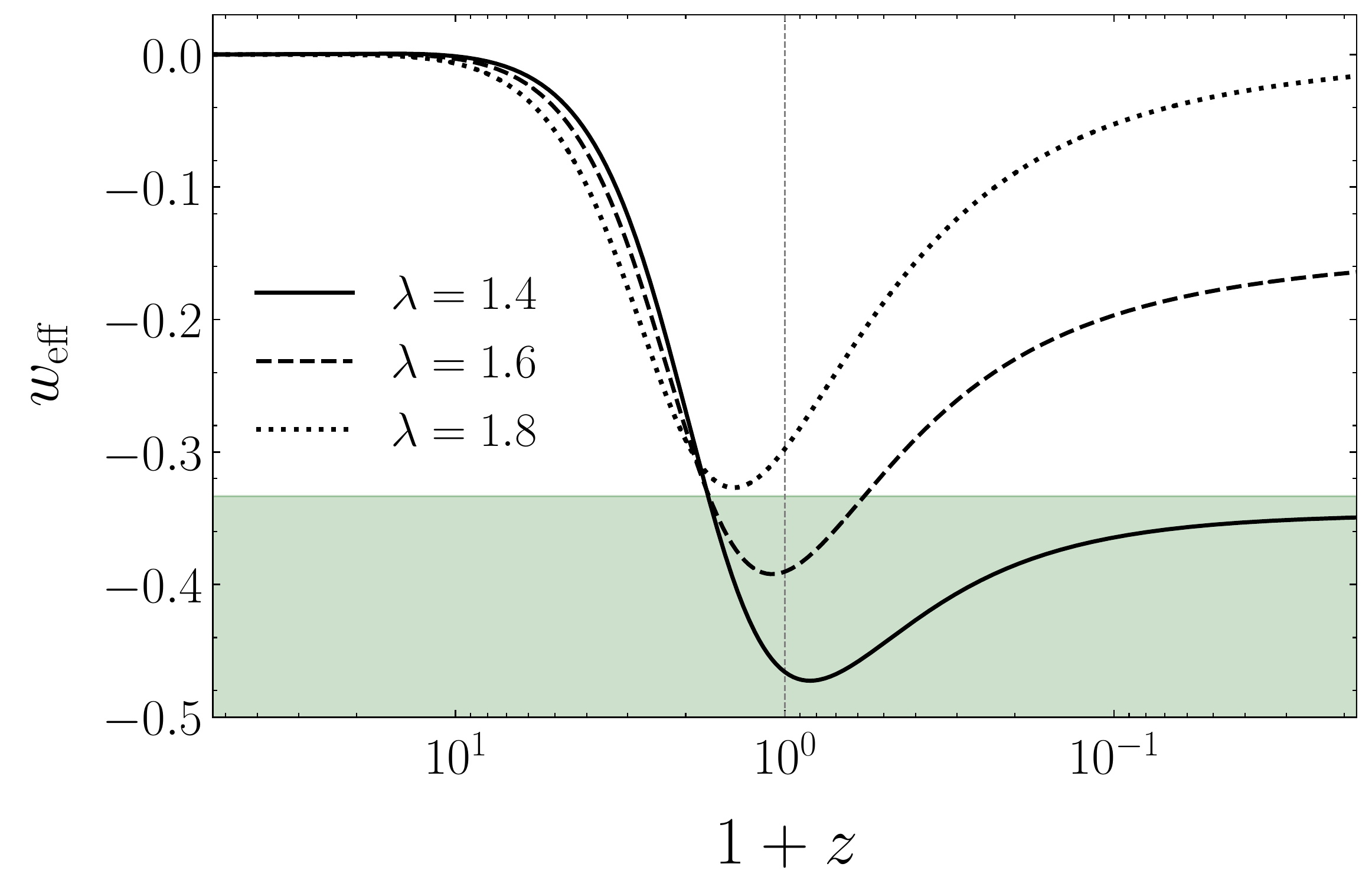}}
  \caption{\label{fig:fig3} \underline{Left Panel}: Evolution of the energy densities of matter (black) and scalar field (pink) for different ICs for the field's velocity, $\phi'_i$, with fixed $\alpha=0.03$ and $\lambda = 0.2$. \underline{Right Panel}: Effective equation of state, $w_{\rm eff}$, for different values of $\lambda$, namely $\lambda=1.4$ (solid line), $\lambda=1.6$ (dashed line) and $\lambda=1.8$ (dotted line), with fixed $\alpha=0.03$. The shaded green area corresponds to the region where the Universe features accelerated expansion, {\it i.e.} ${w_{\rm eff}<-1/3}$. }
\end{figure}

At this point, there is a subtlety that should be noted. Although the ICs for the scalar field do not have any influence on the  parameter constraints, there is a link between the initial values of the velocity of the field and the dark energy density, as expressed in Eq.~\eqref{rho_phi}, which will have a subtle impact on the early behaviour of the quintessence. Increasing the initial density of the field inevitably leads to an earlier onset for the scaling regime  by taking higher values of $\phi'_i$. On the other hand, the value for the initial velocity is completely negligible when it comes to setting the time for which the field exits the scaling and starts evolving towards the accelerating attractor. This implies that the duration of the period in which the energy density of dark energy scales with matter is extended for increasing values of $\phi'_i$. This trend is illustrated in the left panel of Fig.~\ref{fig:fig3}. Nonetheless this does not mean that $\phi'_i$ can take any arbitrary value, as the conditions in Eq.~\eqref{scalar_condition} still have to be verified, in order to avoid instabilities in the theory. On the other hand, the initial value for the field {\it per se} has no influence over the dynamics. Indeed $\phi_i$ only appears in the exponential term of the potential, Eq.~\eqref{eq:pot}, which can be equivalently absorbed by the shooting parameter $V_0$.

Finally, we conclude by providing some concrete examples to support the argument in Section \ref{Sec:Parspace}, namely that values of $\lambda^2>2$ can still give rise to present time accelerated expansion under exceptional conditions. In the right panel of Fig.~\ref{fig:fig3} we illustrate the behaviour of the effective equation of state parameter close to the present epoch and up to some time in the future for different values of $\lambda$, and for a fixed coupling parameter, ${\alpha=0.03}$. Indeed we notice  that transient acceleration phases around the present time are achieved for ${\lambda^2 >2}$, before crossing the boundary given by $w_{\rm eff} < - 1/3$, and exiting this region at some point in the future. Accordingly, these solutions may still stand as cosmologically valid, and such values for $\lambda$ need to be taken into account in the statistical analysis of Sec.~\ref{sec:obs}.

\begin{figure*}[b!]
      \subfloat{\includegraphics[height=0.5\linewidth]{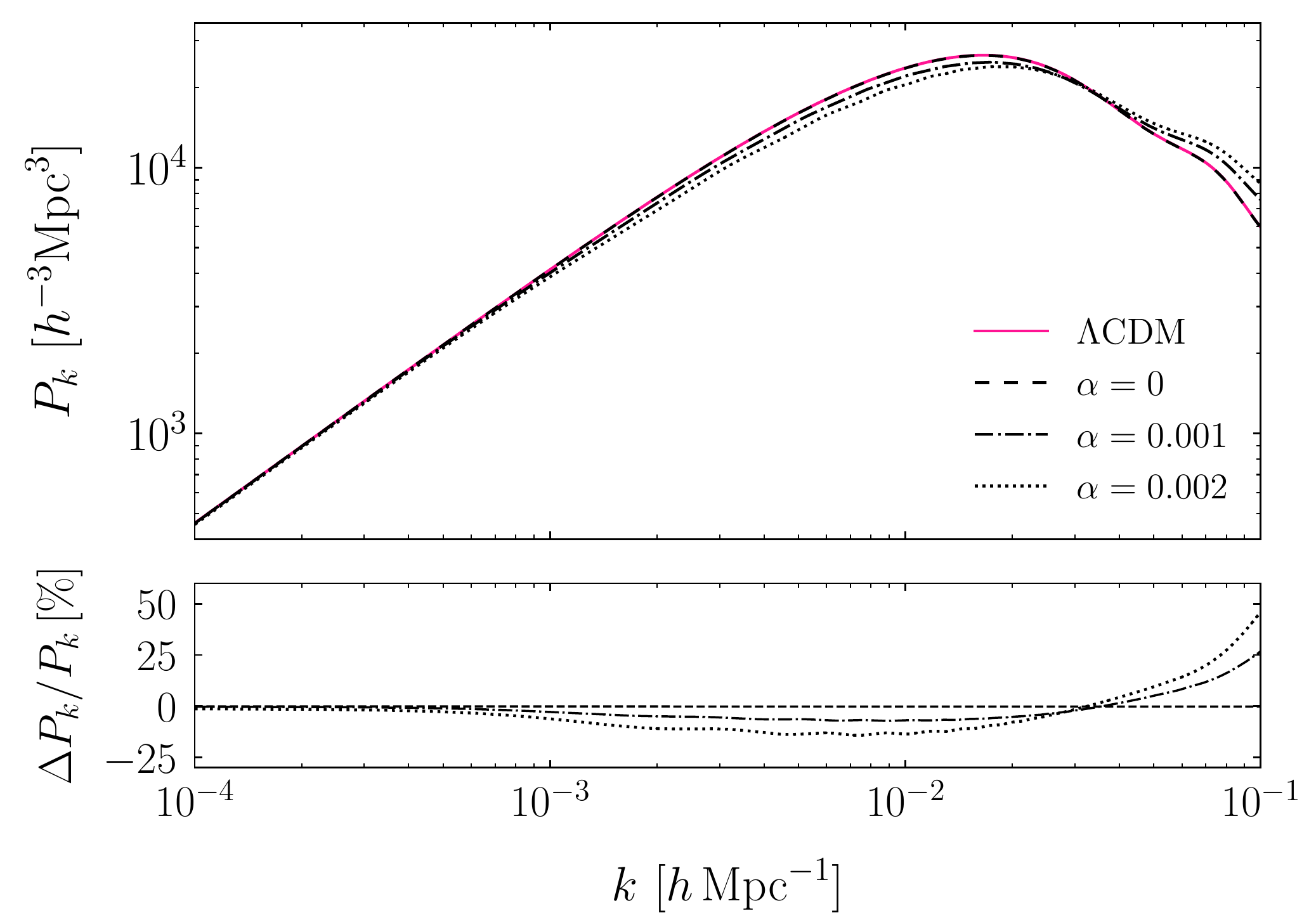}}
  \caption{\label{power_spectrum} \underline{Upper panel}: The matter power spectrum as function of $k$, for the uncoupled case (dashed line), ${\alpha = 0.001}$ (dot-dashed line), ${\alpha = 0.002}$ (dotted line) and $\Lambda$CDM (pink solid line). \underline{Lower panel}: Percentage deviations of the matter power spectrum of the Kinetic model and the uncoupled case from the $\Lambda$CDM model.}
\end{figure*}

\begin{figure*}[h!]
      \subfloat{\includegraphics[width=\linewidth]{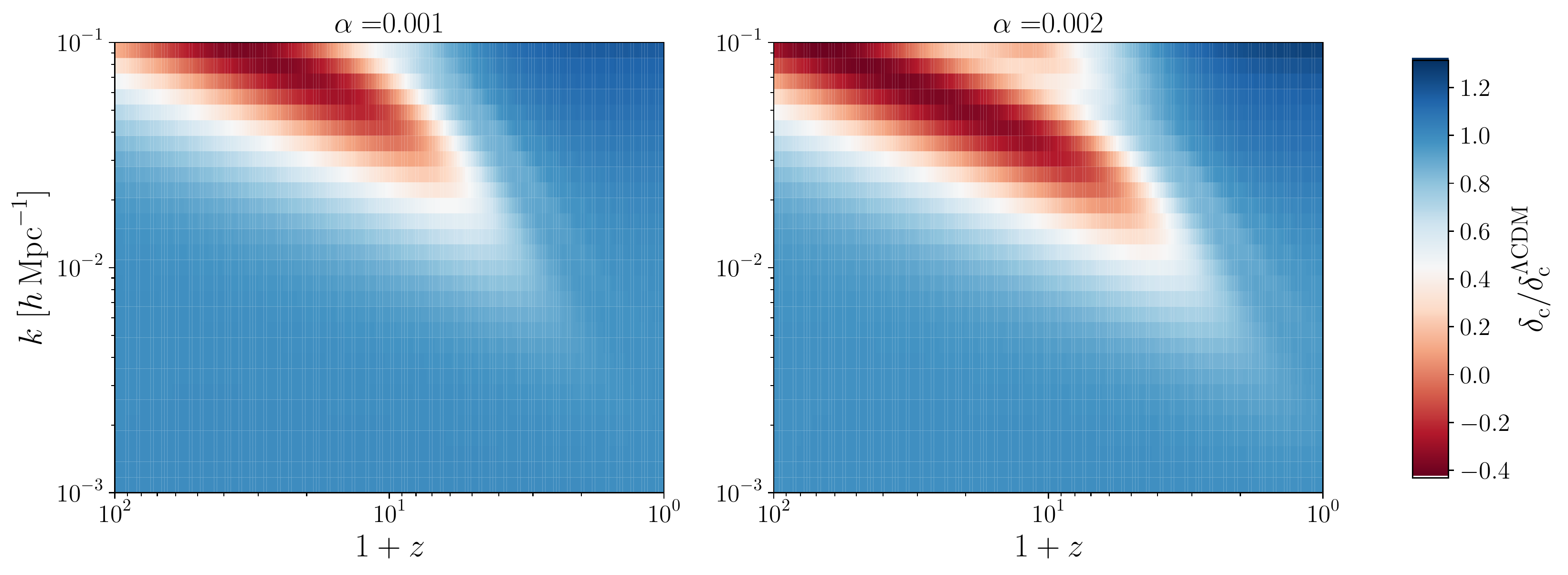}}
  \caption{\label{fig:growth} Evolution of the density contrast of cold dark matter for the Kinetic model relative to the $\Lambda$CDM case, that is $\delta_c/\delta^{\Lambda CDM}_c$, as a function of the Fourier scale $k$ and the redshift $z$ for $\alpha = 0.001$ and $\alpha = 0.002$.}
\end{figure*}

%%%%%%%%%%%%%%%%%%%%%%%%%%%%%%%%%%%%%%%%%%%%%%%%%%%%%%%%%%%%%%%%%%%%%%
\subsection{Cosmological observables}\label{sec:cosmoobs}

In this section we discuss the effect of the coupling on some relevant cosmological observables such as the matter power spectrum and the CMB temperature-temperature (TT) and lensing angular power spectra. We assume adiabatic perturbative initial conditions with an amplitude of curvature fluctuations of $A_s=2.215\times 10^{-9}$, at the pivot scale $k_{\rm piv}=0.05$ Mpc$^{-1}$, and with the spectral index set to $n_s = 0.962$ \cite{Planck:2018vyg}. The remaining cosmological parameters and $\lambda$ are the same as used in the previous Section. We adopt a different set of values for $\alpha$, which are one order of magnitude smaller than the ones used in the numerical analysis of the background quantities, with the reason being that the latter would lead to drastic effects on the cosmological observables. On the contrary, the values we will use to highlight the features on cosmological observables do not produce any significant effects on the background quantities.  As a consequence the features we will show in this Section are attributed solely to the modifications to the linear perturbation equations presented in Section \ref{sec:prts}. Then, for illustrative purposes, we set $\alpha$ to be $1\times 10^{-3}$ and $2\times 10^{-3}$. Moreover, and without loss of generality, we assume vanishing ICs for the scalar field perturbation and its velocity, that is, $\delta\phi (z_i) = \delta\phi' (z_i) = 0$, respectively. 

In the upper panel of Fig.~\ref{power_spectrum} we present the linear matter power spectrum  at present time up to the scale $k_{\rm max}= 0.1 h$ Mpc$^{-1}$, above which the linear perturbative approximation is expected  to break down due to non-linear effects, dominant at smaller scales. In the lower panel we also plot the fractional differences between the coupled scenarios and the $\Lambda$CDM one. We note that the matter power spectrum of the Kinetic model is significantly suppressed at intermediate scales, $10^{-3} h$ Mpc$^{-1}\lesssim k \lesssim 3\times 10^{-2} h$ Mpc$^{-1}$,  with respect to $\Lambda$CDM, and enhanced at the smaller scales. These signatures emerge as a combination of the effects produced by the changes in the  evolution of the background and the cold dark matter perturbations due to the positive exchange of energy that flows from cold dark matter to dark energy. Because the radiation-matter equality era is shifted towards earlier times, when compared with the uncoupled case (see left panel of Fig.~\ref{fig:fig2}), the turnover in the matter power spectrum is shifted to higher $k$. The growth of the matter perturbations is suppressed at intermediate scales, with deviations from $\Lambda$CDM of $\sim 7\%$ and $\sim 14\%$ for $\alpha = 0.001$ and $\alpha = 0.002$, respectively, and enhanced at the smaller scales, with deviations that can reach  $\sim 45\%$ for $\alpha=0.002$. This is illustrated in Fig.~\ref{fig:growth}, where we can clearly see that the largest deviations occur for scales $0.01h <k<0.1h $ Mpc$^{-1}$ and at large redshift, with some milder modifications close to present time as well for $k \sim 0.1h$ Mpc$^{-1}$. At larger scales $k \sim 0.01h$ Mpc$^{-1}$ the deviations are more accentuated at intermediate redshifts ($z \sim 10$). The plots also show that as expected the largest deviations are present for the higher values of $\alpha$. As a consequence, the value of the amplitude of the matter power spectrum  at present time and scale of 8 $h^{-1}$Mpc, denoted by $\sigma_8$, is expected to be larger for the Kinetic model.

\begin{figure}[t!]
      \subfloat{\includegraphics[height=0.32\linewidth]{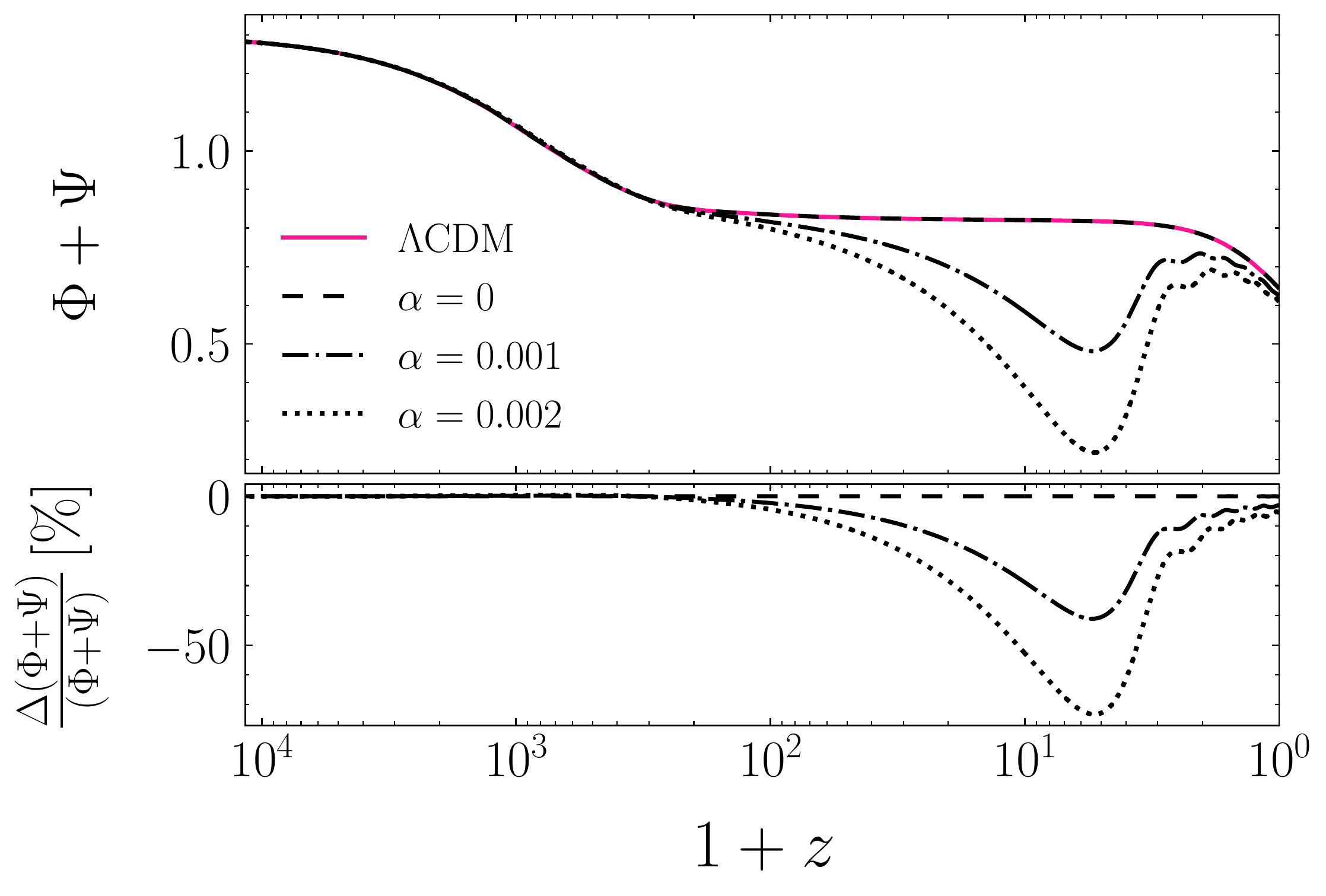}}
      \qquad
      \subfloat{\includegraphics[height=0.32\linewidth]{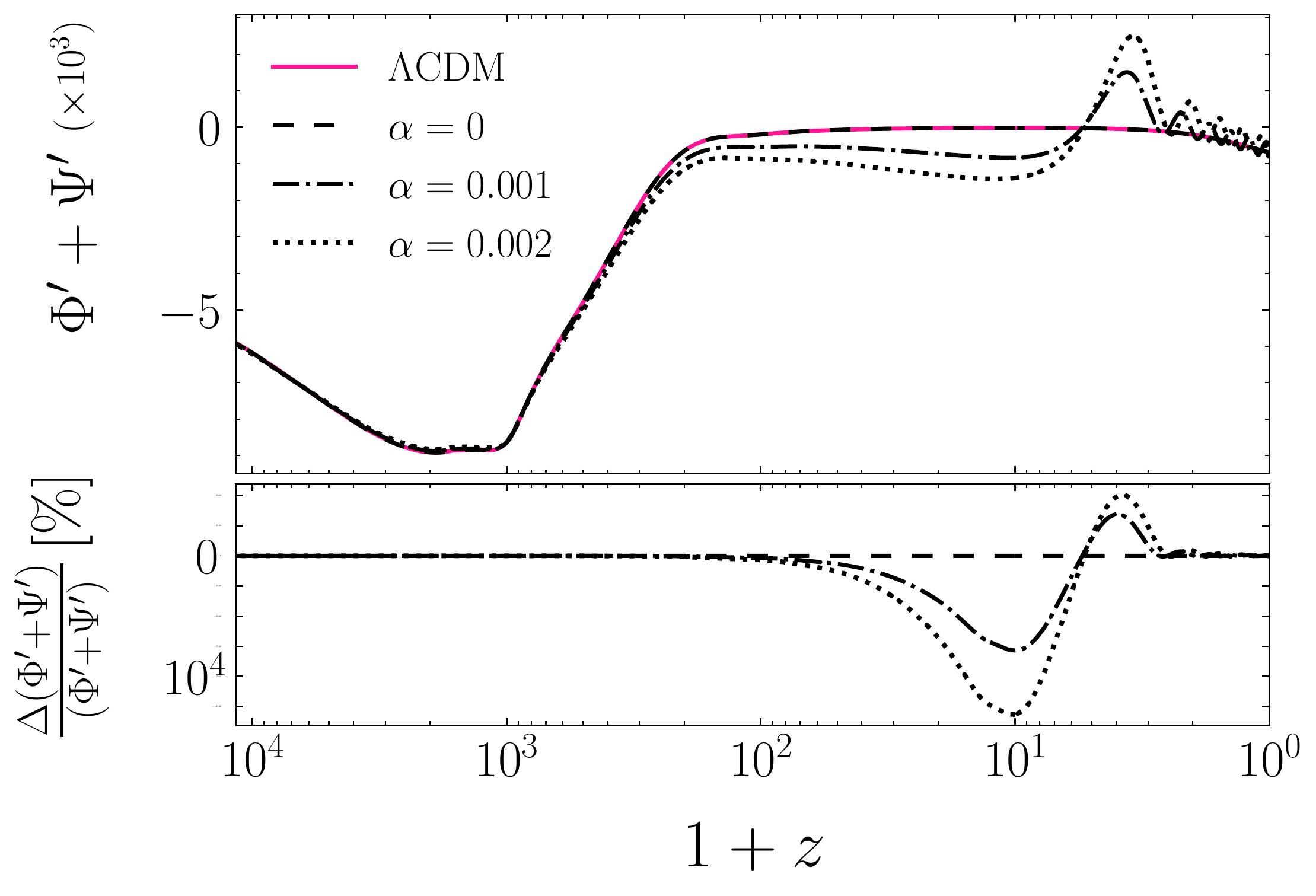}}
  \caption{\label{fig:phipsi} \underline{Left panel}: (Top) Evolution of the sum of the gravitational potentials as a function of the redshift at $k=0.01$ Mpc$^{-1}$ for the cases: uncoupled model (dashed line), $\alpha=0.001$ (dot-dashed line), $\alpha=0.002$ (dotted line), and $\Lambda$CDM (pink solid line). (Bottom) Relative percentage difference of $\Psi+\Phi$ computed with respect to $\Lambda$CDM. \underline{Right panel}: (Top) Evolution of the time derivative of the sum of the gravitational potentials as  a function of the redshift (Bottom) Relative percentage difference of $\Psi^\prime+\Phi^\prime$ computed with respect to $\Lambda$CDM.}
\end{figure}

\begin{figure}[h!]
      \subfloat{\includegraphics[height=0.5\linewidth]{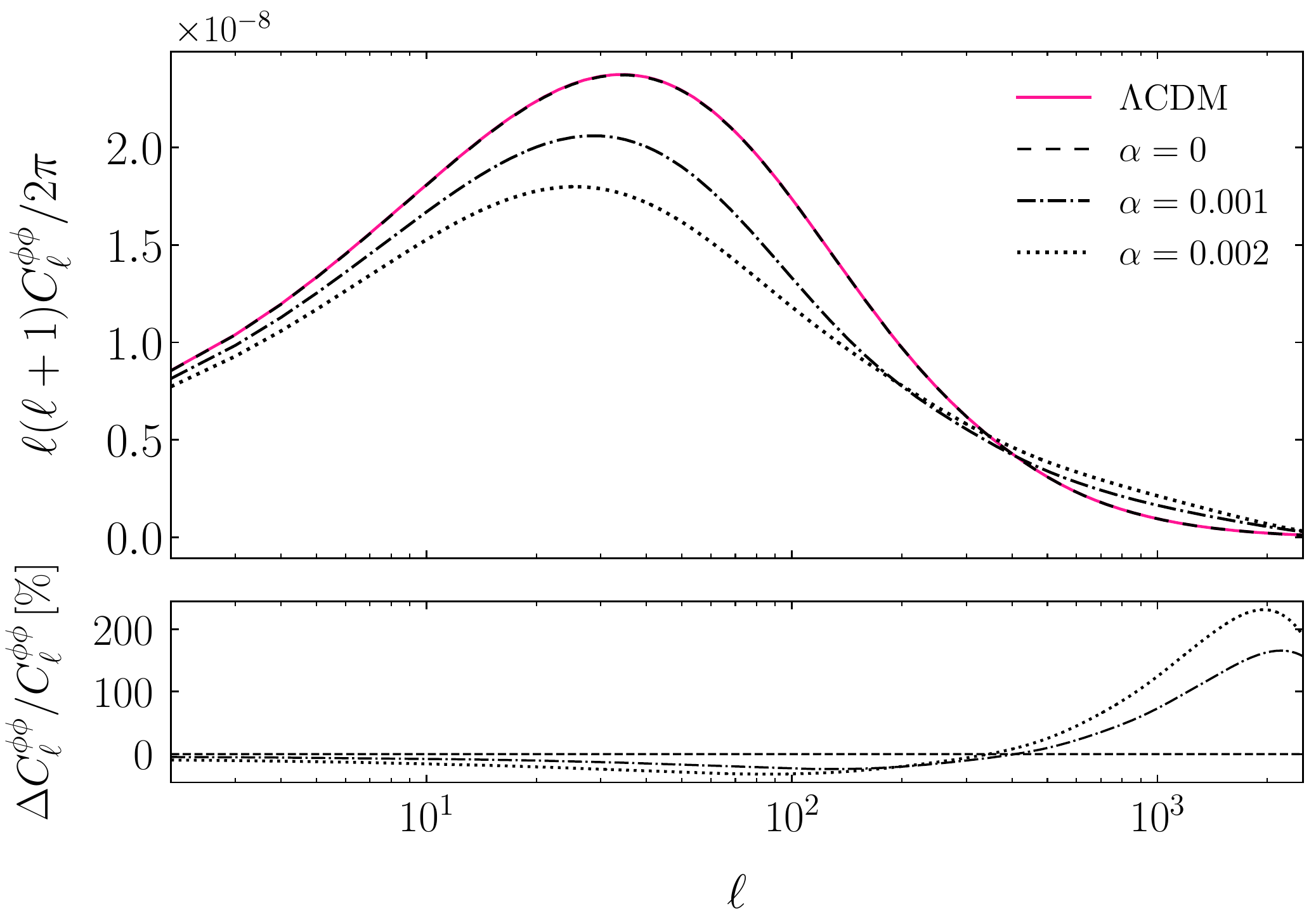}}
  \caption{\label{fig:lensing} 
  \underline{Upper panel}: Lensing angular power spectra for $\Lambda$CDM (solid pink line), $\alpha=0.001$ (dot-dashed line), $\alpha=0.002$ (dotted line) and the uncoupled case (dashed line). \underline{Lower panel}: Relative difference between the lensing power spectra of each model and that of $\Lambda$CDM.}
\end{figure}

\begin{figure}[h!]
      \subfloat{\includegraphics[height=0.5\linewidth]{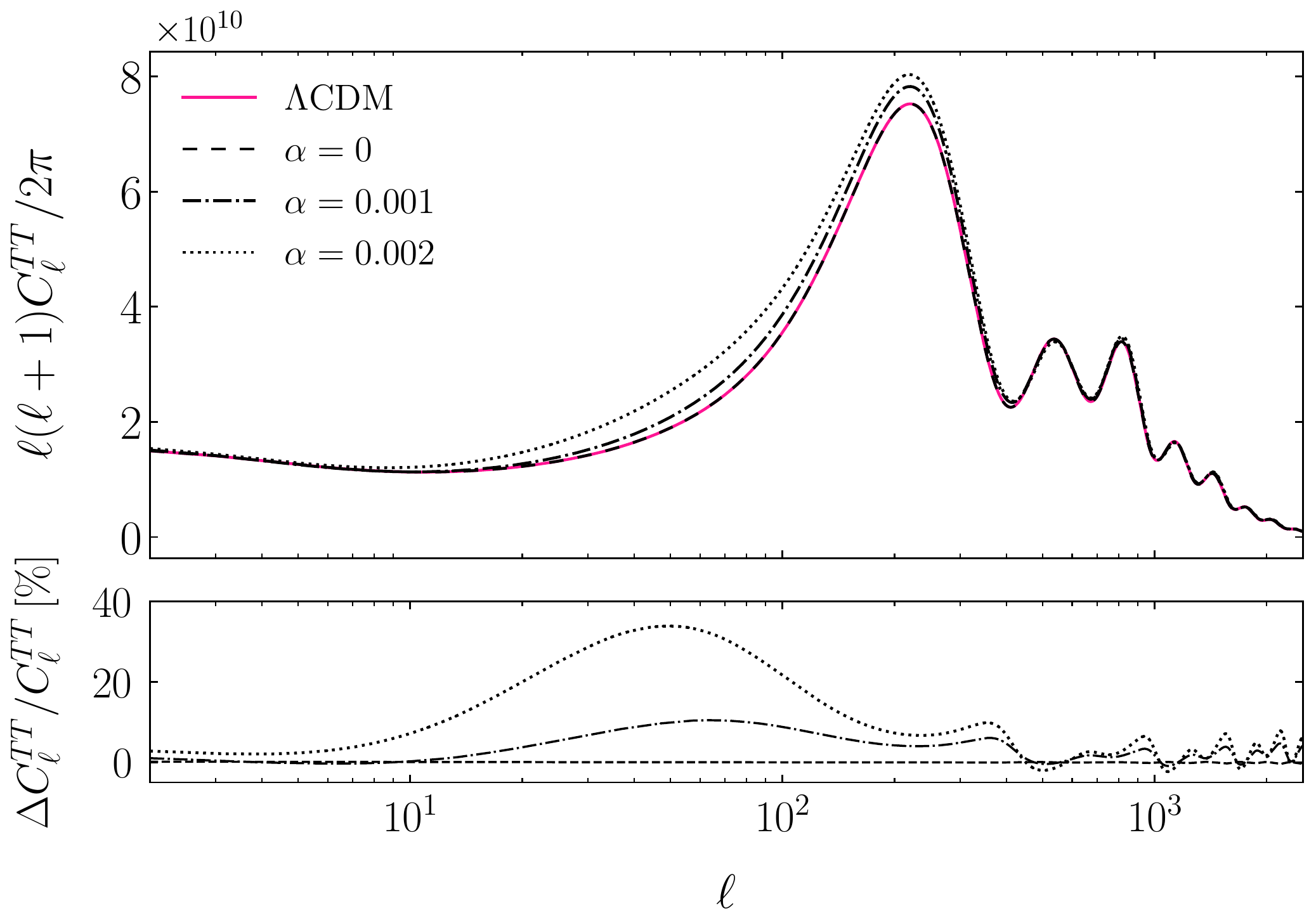}}
  \caption{\label{cmb} \underline{Upper panel}: TT power spectrum as function of the angular scale $\ell$, for the uncoupled case (dashed line), ${\alpha = 0.001}$ (dot-dashed line), ${\alpha = 0.002}$ (dotted line) and $\Lambda$CDM (pink solid line) for reference. \underline{Lower panel}: Percentage deviations of the TT power spectra for the coupled and uncoupled cases with respect to $\Lambda$CDM.}
\end{figure}

In Fig.~\ref{fig:phipsi} we show the sum of the gravitational potentials $\Phi+\Psi$ (left panel) and their time derivative (right panel), as a function of the redshift, for a fixed scale, $k=0.01$ Mpc$^{-1}$. The evolution of the potentials is regulated according to the Poisson equation. We can infer that the value of the lensing potential, given by $\phi_{\text{lens}}=(\Psi+\Phi)/2$, is lower in the Kinetic model when compared to the standard cosmological scenario, resulting in a suppression of the lensing power spectrum, as shown in Fig.~\ref{fig:lensing}. This effect becomes increasingly evident for larger value of $\alpha$. The quantity $\Psi^\prime+\Phi^\prime$ instead is directly connected with the integrated Sachs-Wolfe effect (ISW). The latter affects the shape of the TT power spectrum as it enters in  the radiation transfer function. The total ISW effect is divided into: an early time contribution, produced during the transition from radiation to matter dominated epochs, which in the Kinetic model is shifted towards earlier times when compared to the standard scenario; and a late time contribution, related with the presence  of the dark energy component. The impact of the ISW effect on the TT power spectrum is illustrated in Fig.~\ref{cmb}, as a function of the angular multipole $\ell$, exhibiting an overall enhancement with respect to the reference case for $\ell \lesssim 300$. While milder differences are identified around the \textit{plateau} at $\ell <10$, significant deviations can be appreciated around $10< \ell < 200$, in particular for $\ell\sim 50$, being as large as $\sim40\%$ for $\alpha = 0.002$. Moreover, there is a clear increase in the amplitude of the first peak, accompanied by a broadening of its shape. Likewise, the presence of the coupling and the modifications to the background expansion also induce small differences between the peaks and troughs at the higher multipoles. 

These effects can be measured using cosmological data from background and large-scale structure.

%%%%%%%%%%%%%%%%%%%%%%%%%%%%%%%%%%%%%%%%%%%%%%%%%%%%%%%%%%%%%%%%%%%%%%

\section{Cosmological constraints and model selection analysis}\label{sec:obs}

In this Section we present the constraints on the cosmological and model parameters of the Kinetic model for different combinations of data sets. We perform a Bayesian Monte Carlo Markov Chain (MCMC) analysis using the Metropolis-Hastings algorithm 
implemented in the \texttt{Monte Python}\footnote{\href{https://github.com/brinckmann/montepython_public}{https://github.com/brinckmann/montepython\_public}} sampler \cite{Audren_2013,Brinckmann:2018cvx} interfaced with our personal modified version of \texttt{CLASS}\footnote{\href{https://github.com/lesgourg/class_public}{https://github.com/lesgourg/class\_public}} \cite{lesgourgues2011cosmic,Blas_2011,lesgourgues2011cosmic2}. The general aim is to estimate the sample posteriors that maximise the likelihood associated to each data set, therefore minimising the statistical error distribution. Subsequently, we analyse the MCMC chains and produce the results reported in Tables~\ref{tab:Kineticbounds} and \ref{tab:LCDMbounds}, and in Figures~\ref{fig:const_kinetic}, \ref{fig:1D_kinetic_lcdm}, \ref{fig:OmH0S8}, and \ref{fig:Om_lambda}, resorting to the \texttt{GetDist}\footnote{\href{https://github.com/cmbant/getdist}{https://github.com/cmbant/getdist}} Python package \cite{Lewis:2019xzd}. For comparison purposes we also report on the constraints derived for the standard cosmological scenario. Finally we examine whether the Kinetic model is supported by the data over $\Lambda$CDM. 

%------------------
\begin{table}[h!]
\begin{center}
\begin{tabular}{c|c}
\hline
Parameter                    & Prior \\
\hline
$\Omega_b h^2$                & $[0.005,0.1]$ \\
$\Omega_c h^2$                & $[0.001,0.99]$ \\
$100\theta_{s}$                    & $[0.5,10]$ \\
$z_{reio}$                          & $[0.,20.]$ \\
$n_s$                      & $[0.7,1.3]$ \\
$\log \left(10^{10}A_{s} \right)$   & $[1.7, 5.0]$ \\
\hline
$\lambda$                           & $[0,2]$ \\
$\alpha$                            & $[0,1]$ \\
\hline %S
\end{tabular}
\end{center}
\caption{Flat priors on the cosmological and model parameters sampled in this work.}
\label{tab:priors}
\end{table}
%------------------

%----------------------------------------
\begin{table}[]
    \centering
    \begin{tabular}{c|c|c|c}
     \hline
     \multicolumn{4}{c}{Kinetic Model}\\
    \hline
Parameter          &  Plk18          & Plk18+BAO+SN & Plk18+BAO+SN+len\\
\hline
$S^0_8$            & $0.793^{+0.110}_{-0.064}$             & $0.875^{+0.037}_{-0.043}$ & $0.863^{+0.030}_{-0.039}$ \\
$\Omega^0_m$ & $0.257^{+0.045}_{-0.025}$           & $0.2988^{+0.0072}_{-0.0036}$ & $0.2982^{+0.0070}_{-0.0035}$ \\
$H_0$              & $64.0^{+3.3}_{-1.8}$              & $67.14\pm 0.62$ & $66.94^{+0.60}_{-0.54}$ \\
$10^{-9}A_{s }$              & $2.088\pm 0.035$           & $2.096\pm 0.035$ & $2.111\pm 0.031$ \\
$n_s$              & $0.9667\pm 0.0047$           & $0.9669\pm 0.0044$ & $0.9655\pm 0.0041$ \\
\hline
$\lambda$                & $1.11\pm 0.48$            & $0.42^{+0.18}_{-0.21}$  & $0.41^{+0.17}_{-0.22}$\\
$10^{4} \alpha$          & $1.88\pm 0.95$     & $1.37^{+0.67}_{-1.00}$ & $1.05^{+0.51}_{-0.87}$ \\
\hline
    \end{tabular}
    \caption{$68\%$ C.L. bounds on the cosmological and model parameters for the Kinetic model for the three different combinations of data sets: \textit{Planck}, \textit{Planck} combined with BAO and SN, and  their full combination with CMB lensing.}
    \label{tab:Kineticbounds}
\end{table}

%-----------------
%-----------------
\begin{table}[t!]
    \centering
    \begin{tabular}{c|c|c|c}
    \hline
     \multicolumn{4}{c}{$\Lambda$CDM Model}\\
    \hline
Parameter          &  Plk18         & Plk18+BAO+SN & Plk18+BAO+SN+len\\
\hline
$S^0_8$            & $0.833\pm 0.016$             & $0.831^{+0.013}_{-0.015}$ & $0.834\pm 0.013$ \\
$\Omega^0_m$ & $0.3163\pm 0.0085$           & $0.3151^{+0.0060}_{-0.0075}$ & $0.3162\pm 0.0073$ \\
$H_0$              & $67.31\pm 0.61$              & $67.39^{+0.53}_{-0.45}$ & $67.32\pm 0.53$ \\
$10^{-9}A_{s }$              & $2.102\pm 0.034$           & $2.102\pm 0.034$ & $2.105^{+0.028}_{-0.032}$ \\
$n_s$              & $0.9652\pm 0.0044$           & $0.9656\pm 0.0039$ & $0.9651\pm 0.0041$ \\
\hline
    \end{tabular}
    \caption{$68\%$ C.L. bounds on the cosmological parameters for the $\Lambda$CDM model for the three different combinations of data sets: {\it Planck} 2018, {\it Planck} 2018 combined with BAO and SN, and and their full combination with CMB lensing.}
    \label{tab:LCDMbounds}
\end{table}

\subsection{Data sets}\label{sec:data}

For the present analysis we resort to the CMB {\it Planck} 2018 \cite{Planck:2019nip} data for large angular scales $\ell =[2,29]$ and a joint of TT, TE and EE likelihoods for the small angular scales. In detail, for the latter case, $\ell =[30,2508]$ for the TT power spectrum and $\ell =[30,1996]$ for the TE cross-correlation and EE power spectra. This will be our baseline data set and we will refer to it as ``Plk18" in what follows. Subsequently, we examine the changes when adding to the Plk18 data set a compilation of BAO distance and expansion rate measurements from the Sloan Digital Sky Survey (SDSS) DR7 Main Galaxy Sample \cite{Ross:2014qpa}, SDSS DR12 consensus release \cite{BOSS:2016hvq} and the 6dF Galaxy Survey \cite{Beutler:2011hx} (see text and Figure~11 in Ref. \cite{Planck:2018vyg} for more details), and distance moduli measurements of type Ia Supernova (SN) data  from Pantheon \cite{Pan-STARRS1:2017jku}, hereafter simply ``Plk18+BAO+SN". Finally we consider the combination of ``Plk18+BAO+SN" with the addition of the CMB lensing potential data from {\it Planck} 2018 ~\cite{Planck:2019nip,Planck:2018lbu}, referenced as ``Plk18+BAO+SN+len'' from now on. We note that both the CMB \textit{Planck} 2018 temperature and polarisation angular power spectra data used corresponds to the \textit{standard} reference likelihood from the 2018 release \footnote{\href{http://pla.esac.esa.int/pla}{http://pla.esac.esa.int/pla}} used in the \textit{Planck} analysis. In particular this is given by the product of the \texttt{Commander}, \texttt{SimALL}, and \texttt{PlikTT,TE,EE} likelihoods \cite{Planck:2019nip}. 

% \begin{table}[]
% \begin{center}
% \begin{tabular}{c|c|c|c}
% \hline
% Observable & $z$ & BAO & Reference \\
% \hline
% $r_s/D_v$ & $0.106$ & $0.327 \pm 0.015$ & \cite{Beutler:2011hx} \\
% $D_V/r_s$ & $0.15$ & $4.47 \pm 0.16$ & \cite{Ross:2014qpa} \\
% $D_M (r_{\rm d,fid}/r_{\rm d})\ ({\rm Mpc})$ & $0.38$ & $1512.39 \pm 24.99$ & \cite{BOSS:2016hvq} \\
% $D_M (r_{\rm d,fid}/r_{\rm d})\ ({\rm Mpc})$ & $0.51$ & $1975.22 \pm 30.10$ & \cite{BOSS:2016hvq} \\
% $D_M (r_{\rm d,fid}/r_{\rm d})\ ({\rm Mpc})$ & $0.61$ & $2306.68 \pm 37.08$ & \cite{BOSS:2016hvq} \\
% $H (r_{\rm d,fid}/r_{\rm d})\ ({\rm km\, s^{-1}\, Mpc^{-1}})$ & $0.38$ & $81.2087 \pm 2.3683$ & \cite{BOSS:2016hvq} \\
% $H (r_{\rm d,fid}/r_{\rm d})\ ({\rm km\, s^{-1}\, Mpc^{-1}})$ & $0.51$ & $90.9029 \pm 2.3288$ & \cite{BOSS:2016hvq} \\
% $H (r_{\rm d,fid}/r_{\rm d})\ ({\rm km\, s^{-1}\, Mpc^{-1}})$ & $0.61$ & $98.9647 \pm 2.5019$ & \cite{BOSS:2016hvq} \\
% \hline %S
% \end{tabular}
% \end{center}
% \caption{Detail on the BAO data used in this work. Not sure about the errors, check!!!}
% \label{tab:baodata}
% \end{table}

Our set of free parameters consists of the baseline $\Lambda$CDM cosmological parameters, namely: ${\Omega_b h^2}$, ${\Omega_c h^2}$, $A_{\rm s}$ and $n_s$, the angular size of the sound horizon at recombination $\theta_s$ and the reionisation redshift $z_{reio}$; moreover we add the two free parameters associated to the Kinetic model, $\alpha$ and $\lambda$. We impose flat priors for all the parameters sampled and these are specified in Table~\ref{tab:priors}\footnote{We use  a linear sampling for $\alpha$ but we have run chains using also  a logarithmic sampling. Comparing the results we concluded that the C.L. bounds and the marginalised posterior distributions found are in agreement at $1\sigma$ level. This supports the robustness of the results reported on with a flat prior.}. We will then provide derived constraints on $H_0$, and $S^0_8 = \sigma_8^0 \sqrt{\Omega_m^0/0.3}$.

%%%%%%%%%%%%%%%%%%%%%%%%%%%%%%%%%%%%%%%%%%%%%%%%%%
\subsection{Cosmological bounds}\label{sec:results}
%%%%%%%%%%%%%%%%%%%%%%%%%%%%%%%%%%%%%%%%%%%%%%%%%%
%-----------------

%------------------
\begin{figure}[t!]
      \subfloat{\includegraphics[width=1.\linewidth]{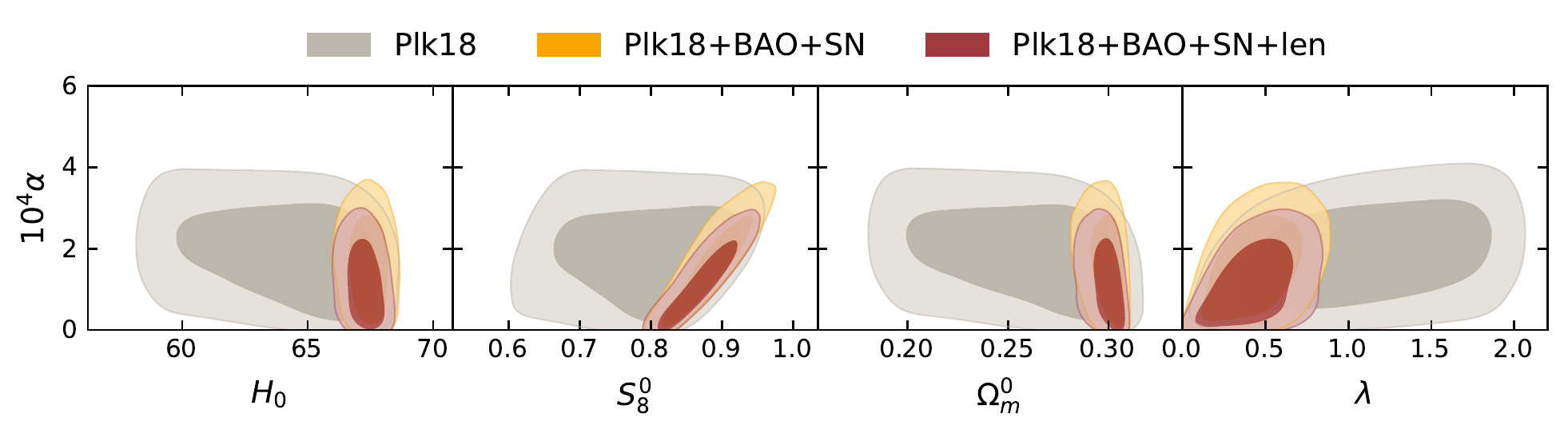}}
  \caption{\label{fig:const_kinetic} $68\%$ and $95\%$ C.L. contours obtained in the Kinetic model under consideration for the {\it Planck} 2018 data (grey), the {\it Planck} 2018, BAO and SN combination (yellow), and their combination with CMB lensing (red).}
\end{figure}
%------------------

%------------------
\begin{figure}[t!]
      \subfloat{\includegraphics[width=1.\linewidth]{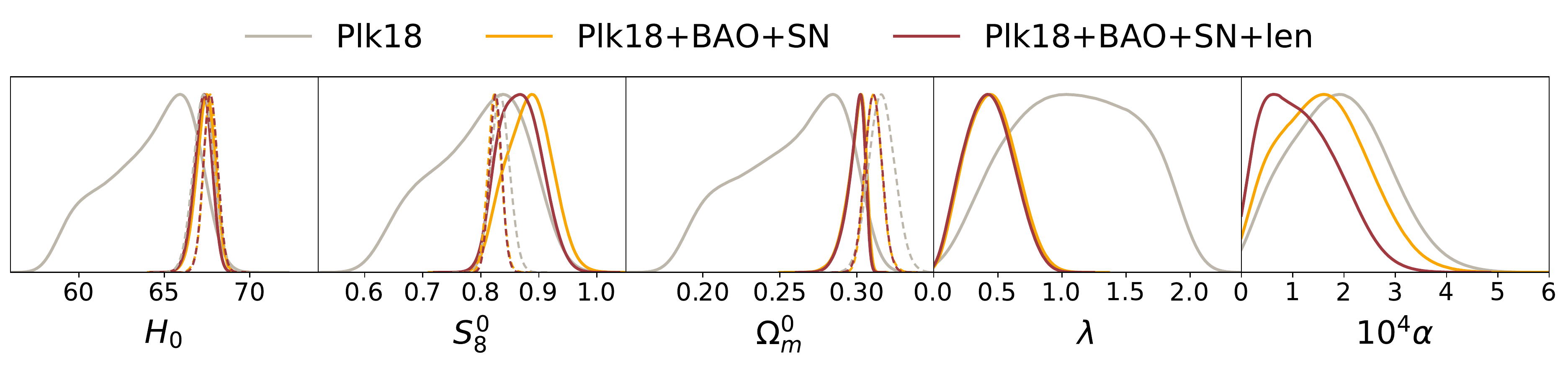}}
  \caption{\label{fig:1D_kinetic_lcdm} 
Comparison between the $\Lambda$CDM (dashed lines) and Kinetic model (solid lines) marginalised likelihood of the  cosmological parameters for the {\it Planck} 2018 data (grey), the {\it Planck} 2018, BAO and SN combination (yellow) and their combination with CMB lensing (red).}
\end{figure}
%------------------

%----------------------------------------

We show the constraints on the model and cosmological parameters for the Kinetic model in Table~\ref{tab:Kineticbounds} and the corresponding contour plots in Figure~\ref{fig:const_kinetic}, for all the data set combinations considered. For comparison purposes, we include the  results for the $\Lambda$CDM model in Table~\ref{tab:LCDMbounds} and in Figure~\ref{fig:1D_kinetic_lcdm}. 

We find that the parameter $\alpha$ is constrained to be of the order of $10^{-4}$, regardless of the combination of data sets considered. The {\it Planck} data alone prefer the higher mean value of $\alpha$, mainly as this allows to better accommodate the TT likelihood; on the other hand, the inclusion of the BAO and SN data results in a slight decrease of the mean  value of $\alpha$; at last, adding the CMB lensing data leads to a shift of the peak of the posterior distribution for the $\alpha$ parameter  to an even lower central  value. This feature is connected to the lensing excess reported by the {\it Planck} collaboration \cite{Planck:2013pxb,Planck:2015mrs,Planck:2018vyg}. As discussed in the previous section, the lensing power spectrum is always suppressed in the Kinetic model, when compared to the $\Lambda$CDM one, with higher values of $\alpha$ corresponding to lower amplitudes of the lensing power spectrum (see Figure~\ref{fig:lensing}). Therefore, in order to better accommodate the CMB lensing data, a lower mean value for $\alpha$ is preferred.

\begin{figure}[t]
    \subfloat{\includegraphics[width=0.75\linewidth]{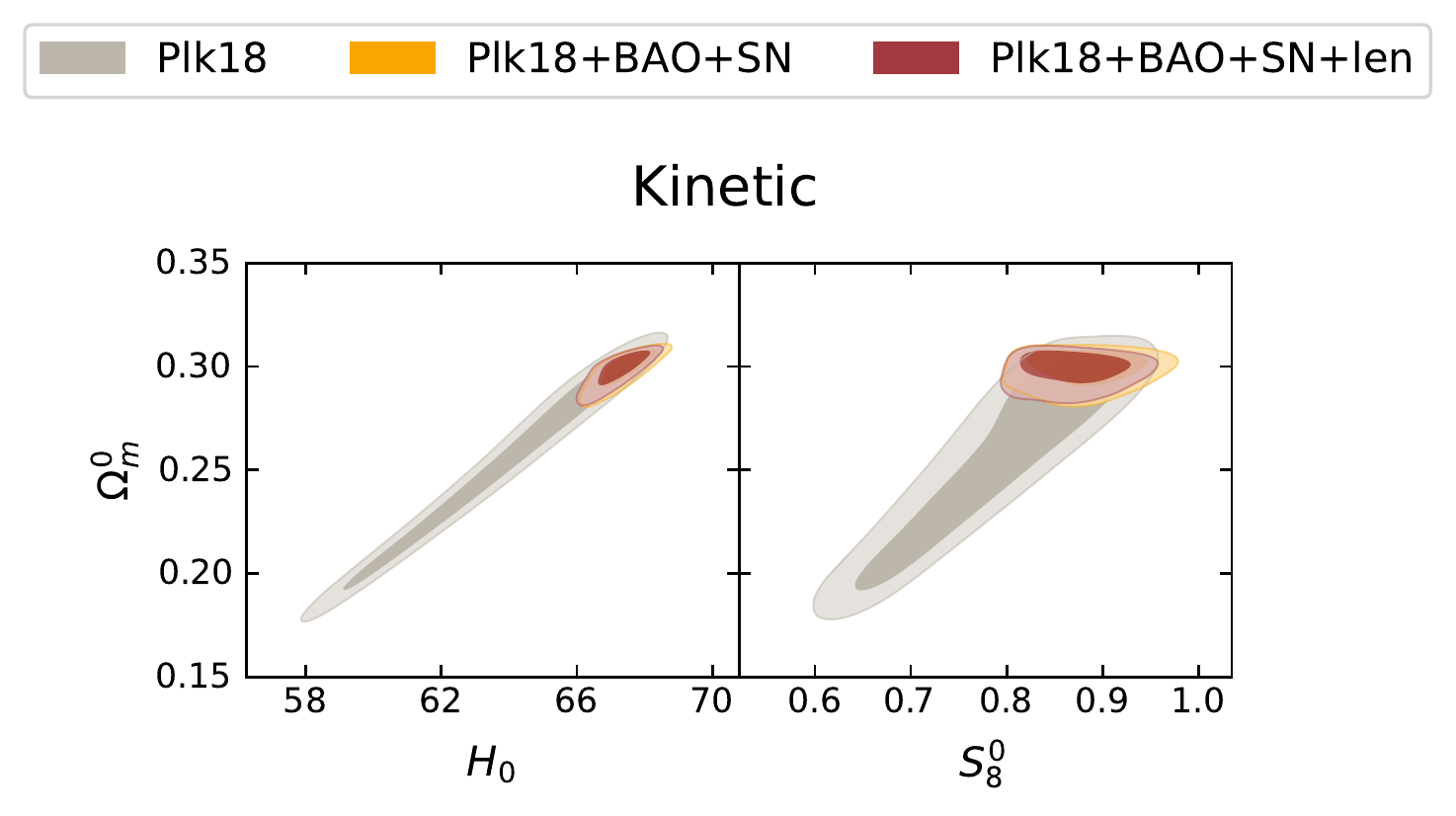}} \\
     \hspace{-5.25pt}\vspace{-15pt}\subfloat{\hspace{-10pt}\includegraphics[width=0.635\linewidth]{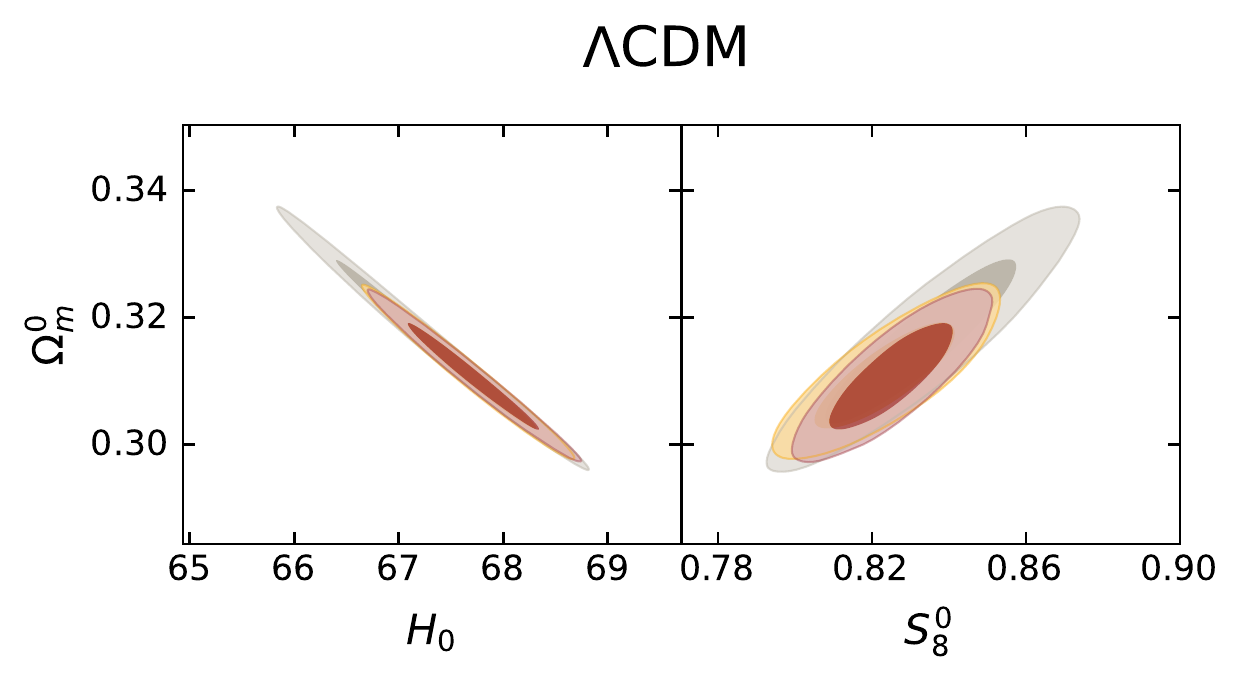}\hspace{10pt}} \\
  \caption{\label{fig:OmH0S8} 68\% and 95\% C.L. 2D contours obtained for the parameters $H_0$ and $\Omega_m^0$ (left panels) and $S_8^0$ and $\Omega_m^0$ (right panels) in the  Kinetic model (upper panels) and $\Lambda$CDM model (lower panels) for the {\it Planck} 2018
data (grey), the {\it Planck} 2018, BAO and SN combination (yellow), and their combination with CMB lensing
(red).}
\end{figure}

Although the constraints on the cosmological parameters of the Kinetic model are compatible with the $\Lambda$CDM ones within the errors, the cosmological standard model yields higher mean values for $H_0$ and $\Omega_m^0$, when compared to the Kinetic model. The latter is characterised by a positive-correlation between $\Omega_m^0$ and $H_0$, contrary to the anti-correlation that characterises the $\Lambda$CDM model, as shown in Figure~\ref{fig:OmH0S8}. In other words, a preference for lower values of $\Omega_m^0$ results in lower values for $H_0$ alike. This characteristic correlation is persistent through all the three data combinations considered. This trait can be ascribed to the presence of a non-vanishing value for the $\alpha$ parameter, associated with an enhancement of the TT power spectrum (see Fig~\ref{cmb}). 

Furthermore, in Figure~\ref{fig:OmH0S8} we depict the contour plots for the constraints in the $S_8^0-\Omega_m^0$ plane. The parameters are positively correlated for both the $\Lambda$CDM model and the Kinetic model. For the latter we find $S_8^0=0.793^{+0.110}_{-0.064}$ at 68\% C.L. with Plk18 data only, thus alleviating the discordance with cosmic shear measurements \cite{Heymans:2020gsg,DiValentino:2020vvd,Abdalla:2022yfr} present in the standard model, for which we report ${S_8^0=0.833\pm 0.016}$. However, as seen in Table~\ref{tab:Kineticbounds} when the other data sets are also taken into account the discrepancy arises again, reflecting a tension between BAO and/or SN data under this framework. A similar situation has also been reported in a Galileon model \cite{Frusciante:2019puu}. This contingency requires further investigation since it has been suggested that there might be a bias towards $\Lambda$CDM-like models enclosed in the BAO data \cite{Carter:2019ulk}. 

\begin{figure}[t!]
      \subfloat{\includegraphics[width=0.5\linewidth]{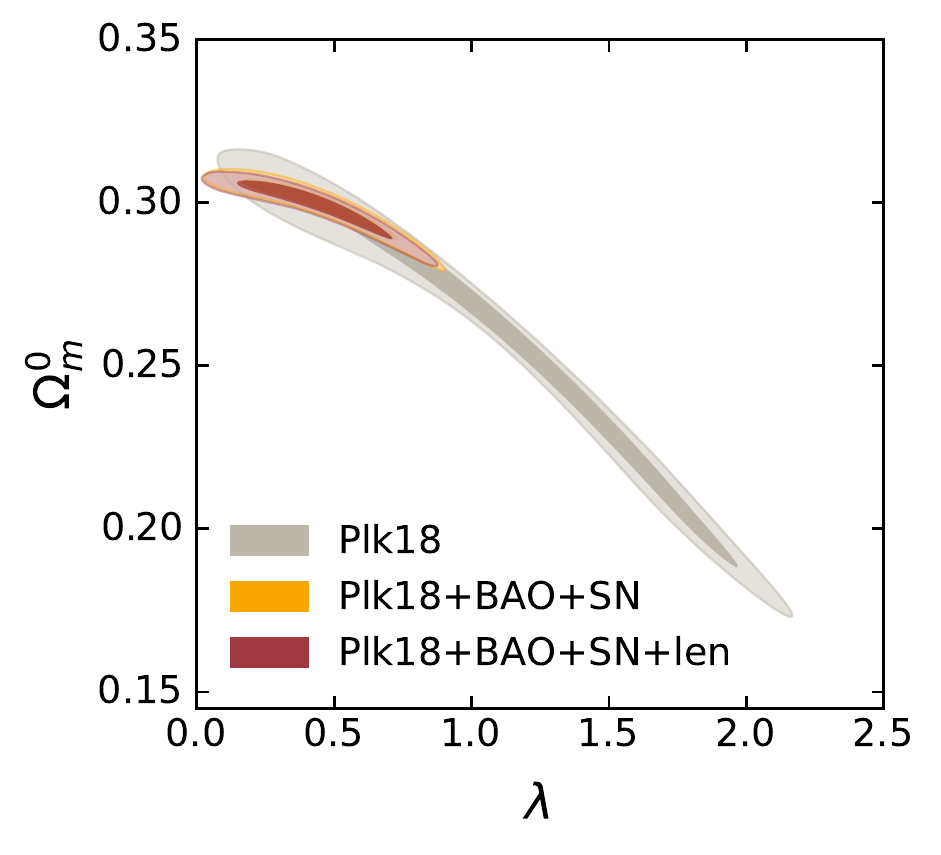}} 
  \caption{\label{fig:Om_lambda} 68\% and 95\% C.L. 2D contours obtained  for the parameters $\lambda$ and $\Omega_m^0$ for the Kinetic model considering the {\it Planck} 2018
data (grey), the {\it Planck} 2018, BAO and SN combination (yellow), and their combination with CMB lensing (red).}
\end{figure}

The inclusion of BAO and SN data leads to narrower constraints on $\Omega_m^0$, which in turn results in tighter constraints on other parameters, such as $H_0$, $S_8^0$, and $\lambda$. The latter is directly connected to the anti-correlation shown in Fig.~\ref{fig:Om_lambda} in the $\Omega_m^0$-$\lambda$ plane, \textit{i.e.} higher values of $\Omega_m^0$ select lower values for $\lambda$. This negative correlation is justified by considering that the late time accelerated expansion, expressed in terms of $w_\phi \approx 1-2V/3H^2$, is mainly regulated by two parameters, namely $\Omega_{\phi}^0$ and $\lambda$. The former is given by the Friedmann constraint $\Omega_{\phi}^0 \approx 1-\Omega_m^0$, meaning that, in turn, higher values of $\Omega_m^0$ are associated with lower values of $\Omega_{\phi}^0$. Therefore, and in order to have a cosmological constant-like scenario for the scalar field at present times, $w_\phi^0\approx -1$, the mean value of $\lambda$ is pushed towards smaller values, explaining the identified anti-correlation between $\Omega_m^0$ and $\lambda$.

\begin{table}[b!]
    \centering
    \begin{tabular}{c|c|c|c}
    \hline
         & \,\,\,\, Plk18\,\,\,\,          & Plk18+BAO+SN & Plk18+BAO+SN+len\\
\hline
$\Delta \chi^2_{\rm eff}$ &  $ -0.9$  & $0.7$ & $1.0$ \\
$\Delta$DIC & $-0.3$  & $0.8$ & $1.6$  \\
\hline
    \end{tabular}
    \caption{Results for the $\Delta \chi^2_{\rm eff}$ and $\Delta \text{DIC}$ obtained as the difference between the Kinetic and $\Lambda$CDM scenarios.}
    \label{Tab:statistics}
\end{table}

Finally, we wish to examine whether the Kinetic model is supported over the $\Lambda$CDM case resorting to statistical indicators: the effective $\chi^2$ corresponding to the maximum likelihood, namely $\chi_\text{eff}^2$, and the Deviance Information Criterion (DIC)~\cite{RSSB:RSSB12062}. The former will enable us to assess whether the Kinetic model is preferred by the data against $\Lambda$CDM, by computing  $\Delta\chi_{\rm eff}^2=\chi^2_{\rm eff, Kinetic}-\chi^2_{\rm eff, \Lambda CDM}$, with a negative outcome standing for a support for it, while a positive result indicates no preference. The DIC will complement this analysis as a tool for quantifying this preference, and it is defined as 
\be
\text{DIC}:= \chi_\text{eff}^2 + 2 p_\text{D},
\ee
where 
${p_\text{D} = \overline{\chi}_\text{eff}^2 - \chi_\text{eff}^2}$, with the upper bar denoting the average of the posterior distribution. According to this definition, the DIC accounts for both the reliability of the fit, through the $\chi_\text{eff}^2$ term, and for the Bayesian complexity of the model, encoded in $p_\text{D}$. Hence, more complex models are disfavoured, in line with a quantitative Occam's razor criteria. Hence, cosmological models with smaller DIC should be preferred over models with larger DIC  \cite{Liddle:2009xe,Peirone:2019aua,Peirone:2019yjs,Frusciante:2019puu,Frusciante:2020gkx,Anagnostopoulos:2021ydo,Rezaei:2021qpq,Albuquerque:2021grl,Atayde:2021pgb}. Finally, the quantity
\be
\Delta \text{DIC} = \text{DIC}_\text{Kinetic} - \text{DIC}_\text{$\Lambda$CDM}\mathcomma
\ee
will indicate support for the Kinetic model over the $\Lambda$CDM scenario provided that $\Delta\text{DIC}<0$. In Tab.~\ref{Tab:statistics} we present the values for both the $\Delta \chi_{\rm eff}^2$ and the $\Delta \text{DIC}$. 
We gather that, by taking the Plk18 data alone, a better fit to the data for the Kinetic model is suggested, compared to the $\Lambda$CDM case, since $\Delta \chi^2=-0.9$. However, when the other data sets are included, this preference is no longer present. This is linked to the fact that the BAO and SN data spoil the fit to the TT likelihood which, after the inclusion of the CMB lensing data, becomes worsened as a result of the Kinetic model predicting a suppressed lensing amplitude, while the CMB lensing data actually shows an excess of power. However, it should be noted that the support of the Kinetic model by the \textit{Planck} data over the standard cosmological scenario is not overly significant ($\Delta {\rm DIC}=-0.3$) and the remaining data combinations indicate a slight preference for the $\Lambda$CDM model. Therefore, we conclude that there is no statistical evidence in support for either of the two models in this analysis.

%%%%%%%%%%%%%%%%%%%%%%%%%%%%%%%%%%%%%%%%%%%%%%%%%%%%%%%%%%%%%%%%%%%%%%

%%%%%%%%%%%%%%%%%%%%%%%%%%%%%%%%%%%%%%%%%%%%%%%%%%%%%%%%%%%%%%%%%%%%%%
\section{Conclusions}\label{sec:conclusions}

In this work we have thoroughly explored  the evolution of the background  and  linear perturbations of the Kinetic model, a coupled quintessence theory characterised by a power-law kinetic interaction, with strength characterised by the parameter $\alpha$. We studied the impact of the coupling between the scalar field and the dark matter fluid on the cosmological observables and we have provided cosmological constraints on the parameters of the theory using CMB, CMB lensing, BAO and SN data. 

We have derived the background and linear scalar perturbation equations and we have modified the public Einstein Boltzmann code \texttt{CLASS}. For our study we have identified the theoretically viable parameter space by enforcing stability requirements such as the absence of ghosts and gradient instabilities. 
These mostly define the range of viability of the parameter $\alpha$. The other additional free parameter of this model is the steepness of the potential function, $\lambda$, which has a crucial role in regulating the late time accelerated expansion. We employed an extended viable range for $\lambda$ compared to what had previously been presented \cite{Barros:2019rdv}, as we allowed for transient accelerated regimes at the present time and not at the future attractor only.  

In Section \ref{Sec:Phenomenology} we studied in detail the phenomenology of the Kinetic model. At the background level we found that a non-vanishing value of $\alpha$ allows for the presence of a scaling regime at early times, during the radiation dominated epoch, according to which the ratio of the densities of the cold dark matter and the scalar field approximately scales with $\alpha$.
The initial condition for the velocity of the scalar field sets how long the quintessence field stays in the scaling regime, hence quantifying the deviations from a cosmological constant behaviour. Furthermore we found that, due to the coupling in the dark sector, energy is being transferred from the dark matter field to the scalar field. We also highlighted the presence of a shift of the radiation matter equality towards earlier times.
These two features have a direct impact on the matter power spectrum: the latter leads to a shift in the position of its peak towards higher $k$ modes, generating in turn a suppression for scales $k\lesssim 3\times 10^{-2}h$ Mpc$^{-1}$, when compared to the $\Lambda$CDM case; the former affects the growth of the matter perturbations on larger $k$, resulting in an enhancement with respect to the standard scenario. Likewise, the differences in the growth of the matter perturbations influence the evolution of the gravitational potentials through the Poisson equation. Consequently we found an overall suppression of the lensing potential (and lensing power spectrum), with respect to $\Lambda$CDM, along with a modified ISW effect which alters the shape of the TT power spectrum for large angular scales. 

These theoretical predictions are then used to provide constraints on the model through a Monte Carlo code for cosmological parameter extraction. We found that the $S_8$ tension is alleviated since $S_8^0=0.793^{+0.110}_{-0.064}$ at 68\% with \textit{Planck} data, while the $H_0$ tension is still present. Regardless of the combination of data considered, the parameter $\alpha$ is consistently constrained to be of the order $10^{-4}$. We also reported on the bounds for the other parameter of the model, $\lambda$, for which the strongest constraints are for the two combinations including BAO and SN data. This is attributed to the strong constraining power of BAO data on $\Omega_m^0$, which indirectly impact the bounds on $\lambda$.
Finally we performed a model selection analysis based on the effective $\chi_{\rm eff}^2$ and Deviance Information Criterion, but we were not able to clearly identify the statistically favoured model between $\Lambda$CDM and the Kinetic model.
We want to stress that the purpose of our work is not to make any claim on the class of models characterised by a kinetic
coupling with cold dark matter but to provide constraints on the parameters of the specific model analysed. The latter being the first tested model
in such class of theories. Actually the present analysis can be considered a starting point to construct and test  new kinetic
coupling models with interesting cosmological signatures.  

In conclusion, we remark that it would  be of interest to consider the Kinetic model for future investigations when new probes from upcoming surveys will be available. This progress will help in shedding light on the tensions and the high accuracy data we expect to collect will allow us  to set a definite preference of one model over the other.

%%%%%%%%%%%%%%%%%%%%%%%%%%%%%%%%%%%%%%%%%%%%%%%%%%%%%%%%%%%%%%%%%%%%%%

\acknowledgments
We thank Eleonora di Valentino for useful comments on the results.
E.M.T. is supported by the grant SFRH/BD/143231/2019 from Funda\c{c}\~ao para a Ci\^encia e a Tecnologia (FCT).  B.J.B. is supported by the South African NRF Grants No. 120390, reference: BSFP190416431035; No. 120396, reference: CSRP190405427545.
N.F. is supported by the Italian Ministry of University and Research (MUR) through the Rita Levi Montalcini project ``Tests of gravity on cosmic scales" with reference PGR19ILFGP. B.J.B., E.M.T. and N.F. also acknowledge the FCT project with ref. number PTDC/FIS-AST/0054/2021. The results of this work were possible thanks to The University of Sheffield’s High Performance Computing (HPC) clusters \texttt{Bessemer} and \texttt{ShARC}.

\appendix

\section{Synchronous gauge}\label{synchronous}

In this Appendix we write the linear perturbations equations of the Kinetic model in synchronous gauge.

We use the following metric to describe perturbations in synchronous gauge
\be
 \dd s^2 = a^2(\tau)\left[ -\dd \tau^2 + \left( \delta_{ij} + h_{ij} \right)\dd x^i \dd x^j \right] \mathcomma
\ee
where the scalar modes of the perturbation components $h_{ij}$ are parameterised in Fourier space,
\be
h_{ij}(\vec{x},\tau)=\int d^3k\, \expe^{\imagi \vec{k}\cdot\vec{x}}\left[ \hat{\vec{k}}_i \cdot \hat{\vec{k}}_j\, h(\vec{k},\tau) + \left( \hat{\vec{k}}_i \cdot \hat{\vec{k}}_j- \frac{1}{3}\delta_{ij} \right)6\eta(\vec{k},\tau) \right] \mathcomma
\ee
with $\vec{k}=k\hat{\vec{k}}$.
 The perturbations in Newtonian gauge are  related with the scalar quantities $\eta$ and $h$ as follows \cite{Ma:1995ey}: 
\begin{eqnarray}
\Psi &=& \frac{1}{2k^2}\left[ h'' +6 \eta'' + \mathcal{H}\left( h'+6 \eta' \right) \right] \mathcomma \\
\Phi &=& \eta - \frac{\mathcal{H}}{2k^2}\left( h' + 6\eta' \right) \mathcomma
\end{eqnarray}
where a prime denotes derivatives with respect to the conformal time $\tau$. We can the write the system of equations (\ref{eq:lEE1})-(\ref{eq:lEE4}) in synchronous gauge as follows:
\begin{eqnarray}
k^2 \eta - \frac{1}{2}\mathcal{H}h' &=& -4\pi G a^2 \sum_i \delta \rho_i \mathcomma \\
k^2\eta' &=& 4\pi G a^2 \sum_i \rho_i (1+w_i) \theta_i \mathcomma  \\
h'' + 2\mathcal{H}h' - 2k^2\eta &=& -24\pi G a^2 \sum_i \delta p_i \mathcomma \\
h'' + 6\eta'' + 2 \mathcal{H}\left( h' + 6\eta' \right) - 2k^2\eta &=& 0 \mathperiod
\end{eqnarray}

Similarly one can find the equivalent of the linear perturbation equation for the matter density perturbations and velocity:
\begin{eqnarray}
\delta'_i+3\mathcal{H}\left( \frac{\delta p_i}{\delta \rho_i}-w_i \right)\delta_i + (1+w_i)\left( \theta_i + \frac{h'}{2} \right) &=& 0 \mathcomma \\
\theta_i' +\left[ \mathcal{H}(1-3w_i)+\frac{w_i'}{1+w_i} \right]\theta_i-\frac{\delta p_i}{\delta \rho_i}\frac{k^2}{1+w_i}\delta_i &=& 0 \mathcomma
\end{eqnarray}
and for the cold dark matter density and velocity perturbations:
\be
\delta_c' + \theta_c + \frac{h'}{2} = \frac{Q}{\rho_c} \left( \phi'\delta_c - \delta\phi' \right) - \frac{\phi'}{\rho_c}\delta Q \mathcomma
\ee
\be
\theta'_c + \mathcal{H}\theta_c = \frac{Q}{\rho_c}\left( \phi'\theta_c - k^2\delta\phi \right) \mathcomma \label{eq:cdmtheta2}
\ee
where
\begin{eqnarray}
\delta Q &=& \frac{2\alpha\rho_c}{2\alpha a^2 \rho_c + (1+2\alpha)\phi'^2}\left\{ \frac{h'}{2}\phi' -\phi'\theta_c + \left[ 3\mathcal{H}\phi' + a^2 (V_{,\phi}-Q) \right]\delta_c + \left( 2k^2 + a^2 V_{,\phi\phi} \right)\delta\phi \right. \nonumber \\
&&\left. \hspace{3.8cm} -\left[ 3\mathcal{H}\phi' + 2a^2 (V_{,\phi}-Q) \right] \frac{\delta\phi'}{\phi'} \right\} \mathperiod
\end{eqnarray}
It is worth noting that the synchronous gauge defines a frame which is always comoving with cold dark matter. That is, in the absence of a coupling, $Q=0$, and for an initial condition $\theta_c(z_i)=0$, the the velocity divergence of CDM remains zero throughout time, as dictated by Eq.~\eqref{eq:cdmtheta2}.

Finally we write the equation for the scalar field perturbation:
\be
\delta\phi'' + 2\mathcal{H}\delta\phi' + \left( a^2 V_{,\phi\phi} + k^2 \right)\delta\phi+\frac{h'}{2}\phi' = a^2 \delta Q \mathperiod
\ee

\bibliography{bib}

\end{document}